\documentclass[pdflatex,sn-mathphys-ay]{sn-jnl}

\usepackage{graphicx}%
\usepackage{multirow}%
\usepackage{amsmath,amssymb,amsfonts}%
\usepackage{amsthm}%
\usepackage{mathrsfs}%
\usepackage[title]{appendix}%
\usepackage{xcolor}%
\usepackage{textcomp}%
\usepackage{manyfoot}%
\usepackage{booktabs}%
\usepackage{algorithm}%
\usepackage{algorithmicx}%
\usepackage{algpseudocode}%
\usepackage{listings}%
\usepackage{bm}%
\usepackage{svg}%
\usepackage{placeins}%

\theoremstyle{thmstyleone}%
\theoremstyle{thmstyletwo}%
\theoremstyle{thmstylethree}%

\begin{document}

\title[Monocular 3D bubble reconstruction]{Monocular 3D bubble reconstruction: DNS-informed ray tracing and deep learning}

\author[1,2]{\fnm{Chaitanya S.} \sur{Nayak}}
\equalcont{These authors contributed equally to this work.}

\author[1,2]{\fnm{Faizaan} \sur{Mohammed}}
\equalcont{These authors contributed equally to this work.}

\author[1,2]{\fnm{Vivek} \sur{Kumar}}

\author[1,2]{\fnm{Shivam} \sur{Prajapati}}

\author[1,2]{\fnm{Suhas S.} \sur{Jain}}

\author*[1,2]{\fnm{Cyrus K.} \sur{Aidun}}\email{cyrus.aidun@me.gatech.edu}

\affil[1]{\orgdiv{George W. Woodruff School of Mechanical Engineering}, \orgname{Georgia Institute of Technology}, \orgaddress{\city{Atlanta}, \postcode{30332}, \state{GA}, \country{USA}}}

\affil[2]{\orgdiv{Center for Multiphase Flow Research}, \orgname{Georgia Institute of Technology}, \orgaddress{\city{Atlanta}, \postcode{30332}, \state{GA}, \country{USA}}}

\abstract{Bubble size, depth, and spatial distribution govern interfacial area, momentum and heat and mass transfer, and process uniformity in multiphase flows. A single shadowgraphic view cannot separate bubble diameter from focal-plane displacement, because apparent size, edge sharpness, halo extent, and overlap depend jointly on both, and multicamera systems add optical access, calibration, and cost. Here, we develop a single-camera physics-guided framework combining direct numerical simulation (DNS), experimentally validated Monte Carlo ray tracing, instance-level preprocessing, and focus-conditioned multitask deep learning to recover three-dimensional bubble position and physical-equivalent diameter from two-dimensional shadowgraphs.

DNS-resolved bubbles supply labels for camera-like images rendered over multiple focal-plane settings. The ray tracer reproduces shadowgraphic signatures with errors below \(3\%\) (intensity), \(1\%\) (apparent size), \(15\%\) (blur), \(6\%\) (numerical aperture), and \(4\%\) (edge-spread-function shape), and runs \(\sim30\times\) faster than the CPU implementation. The dataset has \(900\) frames and \(87{,}728\) labeled instances, partitioned by DNS timestamp to prevent data leakage and expanded to more than \(10^{5}\) crops by broad-depth controlled-density rendering. A multitask network predicts signed depth and diameter from each crop and focal-plane coordinate, achieving \(\sim97.6\%\) normalized depth accuracy on held-out data; a closed-loop DNS-to-image-to-reconstruction test of \(180\) bubbles gives a diameter error below \(9.71\%\) (\(\sim90.29\%\) accuracy). On experimental images at four radial planes (\(142\)--\(971\) bubbles each), probability-density-function (PDF) overlap is \(68\%\)--\(92\%\) and log-PDF correlation \(0.72\)--\(0.87\). At the focal reference plane (\(11{,}679\) bubbles), overlap is \(\sim86\%\), correlation \(0.93\), and full-distribution, \(D_{99.8}\), and composite errors are below \(7.7\%\), increasing with focal-plane distance and defining a recoverability limit. Without stereoscopic hardware, the framework offers a scalable diagnostic for dense bubbly flows.}


\maketitle

\section{Introduction}
\label{sec:introduction}

Gas bubbles are central to the dynamics and performance of many multiphase-flow systems, including bubble columns, bubble pressure tensiometer, bioreactors, flotation devices, aerated reactors, and turbulent gas--liquid contactors~\citep{clift1978bubbles,kantarci2005bubble,garcia2009bioreactor,kulkarni2007mass,wang2020regulation,kumar2026viscosity}. Their diameter distribution, spatial organization, deformation, residence time, breakup, and coalescence determine interfacial area and therefore influence momentum coupling, turbulent kinetic-energy exchange, mixing, heat and mass transfer, and process uniformity. In turbulent bubbly flows, these quantities are evolving population statistics shaped by the competition between capillary forces, turbulence, collisions, film drainage, and fragmentation~\citep{hinze1955fundamentals,martinez1999breakup,thomas1981bubble,Liao2010,chan2011film}. Reliable access to bubble-resolved size and three-dimensional position is therefore a prerequisite for connecting measured images to physically meaningful population balances, interfacial-area estimates, and turbulence--bubble interaction models.

The need for depth-resolved information is especially important in turbulent bubbly flows, where bubbles can be deformed, non-spherical, and partially overlapping in the projected image~\citep{hinze1955fundamentals,martinez1999breakup}. Their apparent size and shape therefore depend on physical dimensions, deformation, orientation, and position relative to the focal plane. These effects make reconstruction from two-dimensional images more challenging and limit approaches developed or trained only on isolated, idealized spherical bubbles.

Optical imaging is attractive for this purpose because it is non-intrusive, high-speed, and compatible with transparent laboratory facilities. Shadowgraphy is particularly useful in bubbly flows because refractive bubbles produce measurable variations in apparent size, edge sharpness, halo structure, and image intensity. These quantities, however, depend jointly on the physical bubble size and its axial position relative to the focal plane. If $d$ denotes the physical bubble diameter and $z$ the signed distance from the focal plane, then the apparent diameter, blur, halo extent, and intensity distribution depend on both $d$ and $z$, as well as on bubble deformation and orientation. Consequently, a larger bubble located near the focal plane can produce an image similar to that of a smaller bubble located farther from focus. This makes the inverse problem non-unique when only a single image feature is considered. Moreover, bubbles located on opposite sides of the focal plane can exhibit similar levels of defocus, so predicting only $|z|$ is insufficient for three-dimensional localization. The objective is therefore to recover both the physical bubble size and the signed depth $z$, which identifies whether a bubble lies in front of or behind the focal plane. This inverse problem becomes more challenging in turbulent bubbly flows, where bubbles are non-spherical, partially overlapping, and distributed throughout the measurement depth.

Classical bubble-image analysis has relied heavily on thresholding, edge detection, connected components, Hough-type detection, watershed separation, and related hand-crafted pipelines. These tools remain valuable for extracting projected areas and in-focus population statistics, but their performance often depends on illumination, contrast, overlap, particle or bubble shape, and user-selected parameters. Recent learning-based methods have greatly improved the robustness of bubble detection, segmentation, and tracking. Faster-R-CNN-based models and shape-regression networks have been used for bubble detection and characterization~\citep{haas2020_bubcnn,toyama2025_fasterrcnn}; deep networks have been developed for automated bubble mask extraction in complex two-phase flows~\citep{kim2021_universal}; lightweight instance-segmentation architectures now target industrial bubble inspection with small, low-contrast, and adhesive bubbles~\citep{yang2026_yolo_bubble}; and machine-learning-assisted segmentation has improved repeatable bubble identification and tracking in gas--solid fluidized beds~\citep{jiang2024_ml_bubble_fluidized}. Related work in bubble velocimetry further demonstrates the value of convolutional and optical-flow approaches for extracting motion information from challenging bubble images~\citep{Choi2022BubbleVelocimetry}. These advances show that learned visual representations can outperform hand-crafted rules, but most of them still operate on projected image evidence and therefore return two-dimensional masks, centroids, velocities, or apparent sizes without bubble-resolved physical depth or volume-equivalent diameter.

Three-dimensional bubble measurements can be obtained by adding viewing directions through stereoscopic, multi-camera, or tomographic imaging. Deep-learning-assisted 3D tracking has recently enabled reconstruction of deformable bubbles in swarms, and temporal information can further improve cross-view association and trajectory completeness~\citep{Hessenkemper2024_3DBubbleTracking,wang2025_temporal_3d_bubble_tracking}. These approaches require additional cameras, calibration, synchronization, optical access, and cross-view matching. Those requirements can become restrictive in compact facilities, near-wall measurements, dense or optically distorted flows, and industrial configurations where only one optical path is available. A complementary capability is therefore needed: single-camera reconstruction that uses the defocus and grayscale information already present in shadowgraphic images to infer physically meaningful depth and size.

A second challenge is supervision. Experimental images are abundant, but bubble-resolved three-dimensional labels are rarely available for every object in the field of view. This is precisely the setting in which physics-guided synthetic data can be useful. In other inverse optical problems, such as pendant-drop tensiometry, neural networks trained on numerically generated shapes have been used to infer physical parameters from images~\citep{kratz2020_pendant_drop_ml}. For transparent fluids, recent ray-tracing image simulations have shown that high-fidelity optical modeling can reproduce images of complex refractive objects and can be used to diagnose or refine physical models~\citep{kalita2026_ray_tracing_transparent}. These developments motivate a simulation-to-image strategy for bubbly flows, but the training images must preserve the relevant bubble physics: deformation, finite depth of field, overlap, refractive halos, and realistic grayscale signatures.

Direct numerical simulation (DNS) provides a physically grounded source of such supervision. Modern interface-resolving simulations can retain the instantaneous bubble geometry, bubble-center coordinates, and volume-equivalent diameters while resolving deformation, coalescence, and breakup events in turbulent gas--liquid flows~\citep{jain2020,jain2022a,jain2025stationary}. If the DNS state is supplied to a calibrated optical forward model, every rendered image can be paired with known three-dimensional bubble labels. This pairing converts a difficult experimental labeling problem into a controlled inverse-imaging problem: the forward model creates camera-like observations, while the DNS field supplies the physical depth and diameter targets that are hidden during inference.

However, simulation-based supervision is useful only if the synthetic optical evidence resembles the experimental evidence to be inverted. Idealized disks, spheres, or binary masks do not reproduce the coupled influence of refractive halos, edge-spread functions, defocus, grayscale attenuation, and projected overlap. The present work therefore uses Monte Carlo ray tracing to map DNS-resolved bubble scenes into shadowgraphic images over multiple focal-plane settings. The ray-traced images serve as training inputs and are also validated against experimental shadowgraphic signatures so that the inverse model is trained on optical features that have a measured experimental counterpart. This forward-model validation is an essential step because errors in image formation would otherwise be learned by the inverse model as false physical cues.

The inverse model must also preserve the identity of individual bubbles through several transformations. A full-frame image first contains mixed evidence from focused interfaces, diffuse halos, weak defocused structures, and overlapping instances. The manuscript therefore uses a hierarchical preprocessing pipeline: bubble-related evidence is localized, individual instances are separated, and each retained bubble is converted into a standardized focus-conditioned crop. The crop representation keeps the grayscale image, visible support mask, physical support mask, and known focal-plane coordinate. This construction is designed to retain the local optical features that carry depth information while still allowing the model to use the known imaging geometry.

Depth reconstruction over a broad measurement range introduces another difficulty. Rendering all bubbles across a large depth interval can create excessive projected crowding, causing weak far-defocus signatures to overlap with stronger near-focus evidence. Conversely, training only near the focal plane limits the range over which the model can infer signed depth. To address this tradeoff, the present framework uses broad-depth controlled-density rendering. Bubbles are sampled across focal-plane-relative depth strata so that the training distribution contains near-focus and far-defocus examples while preserving clear correspondence between each rendered optical signature and its DNS-resolved label. Timestamp-disjoint partitioning is then used so that alternate views or focal-plane settings of the same DNS state do not leak across training, validation, and test sets.

Despite substantial progress in bubble imaging and machine learning, a central challenge remains: recovering bubble-resolved three-dimensional information from a single shadowgraphic view. In a single image, the effects of bubble size and distance from the focal plane are coupled, making physical diameter and signed depth difficult to determine independently. Supervised learning could provide a route to this inverse problem, but experimental shadowgraphs generally do not provide bubble-resolved three-dimensional ground truth for training. DNS data can supply the required bubble geometry and position labels, but their use introduces an additional requirement: the corresponding synthetic images must reproduce the optical characteristics of real shadowgraphs, including defocus, intensity variation, deformation, and overlap. The key gap is therefore the absence of a framework that connects three-dimensional numerical bubble data to optically realistic images and uses this paired information to infer both bubble size and signed depth from a single experimental view.

The objective of this work is to determine whether DNS-informed, experimentally validated ray tracing can resolve the coupled depth--diameter ambiguity inherent to single-camera shadowgraphy. Specifically, we examine whether a model trained on paired three-dimensional bubble information and synthetic shadowgraphic images can recover the signed focal-plane-relative depth and physical-equivalent diameter of individual bubbles from a single view. We further quantify how reconstruction accuracy changes with distance from the focal plane, where defocus increases and the optical signatures used for inference become progressively less distinct. The framework combines DNS-resolved bubble geometries with a validated optical forward model to generate physically labeled training data, from which an inverse model learns to reconstruct three-dimensional bubble populations from experimental shadowgraphs (Sec.~\ref{sec:inverse_framework}).

The validation strategy follows the structure of this inverse problem. First, a closed-loop DNS-to-image-to-reconstruction test evaluates whether bubble-level depth and diameter information are preserved through the complete sequence from a known three-dimensional DNS field to a ray-traced image and back to a reconstructed bubble population. Second, experimental transfer is assessed by applying the same inference pipeline to shadowgraph images acquired at known focal-plane positions in a turbulent air--water facility. Because experimental bubble-resolved depth labels are unavailable, reconstructed populations are compared with in-focus experimental reference distributions using full PDF overlap, log-PDF correlation, \(D_{32}\), \(D_{99.8}\), and a recoverability analysis as a function of focal-plane distance.

The present study extends the preliminary depth-aware framework of \citet{nayak2026depth} by replacing experimental proxy labels with DNS-resolved physical supervision, rendering physically evolved bubble geometries in place of idealized particles, validating the forward optical model against experimental shadowgraphic signatures, performing joint depth--diameter regression, and evaluating both closed-loop DNS consistency and experimental population statistics. In this way, the work addresses the inverse shadowgraphic-imaging problem: recovering depth-resolved bubble size information from a single camera without requiring stereoscopic or multi-camera measurements.

\section{Dataset Generation}

\subsection{Experimental shadowgraph dataset}

\begin{figure}[!htbp]
    \centering
    \includegraphics[width=0.7\textwidth]{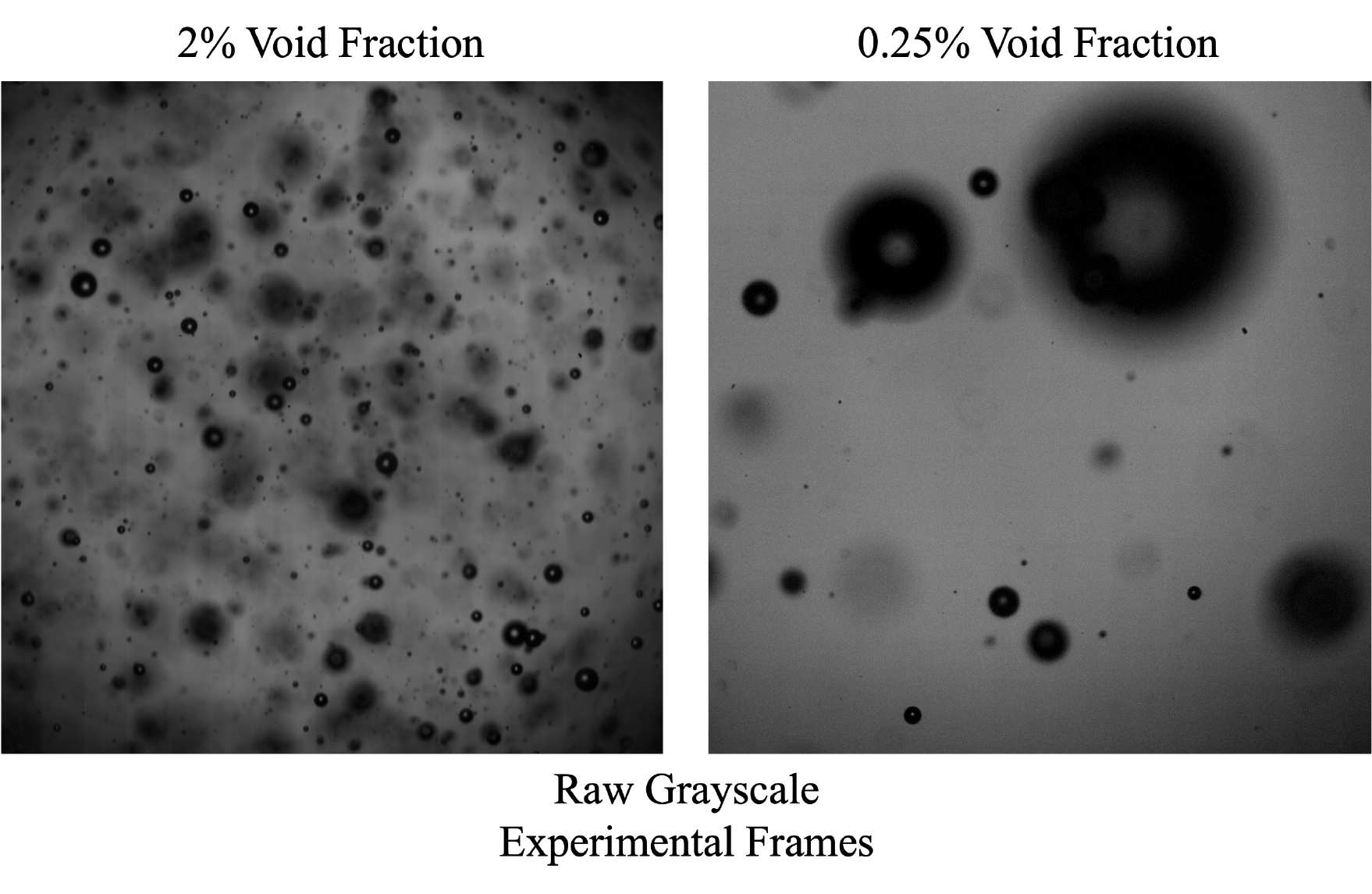}
    \caption{
    Experimental frames containing bubble images at two different void fractions
    }
    \label{fig:data_frames}
\end{figure}

The experimental shadowgraphs used to validate the forward optical model and to assess transfer of the inverse reconstruction framework were acquired in the closed-loop air--water (bubbly flow) facility described by \citet{kumar2026bubble,kumar2026mass}. Air was introduced through an inline sparger upstream of a regenerative turbine pump at void fraction of 0.25\% to 2\%, which dispersed the gas and generated an initially intense turbulent bubbly flow. The mixture subsequently entered a vertical square duct with internal dimensions
\(13.97\pm0.03~\mathrm{mm}\times13.97\pm0.03~\mathrm{mm}\) and a height of
\(609.6\pm3.0~\mathrm{mm}\). The duct was constructed from a rigid polycarbonate frame and optically transparent polycarbonate panels, providing optical access across the channel depth. The pump-generated turbulence was strongest near the duct inlet and decayed rapidly in the streamwise direction; its single- and two-phase characteristics are reported separately by \citet{javadi2026large}. The parent experimental dataset includes measurements at multiple axial locations over \(3.6\leq x/D\leq40\) and at several radial positions spanning the near-wall and centerline regions. Related measurements performed using the same experimental platform. Representative experimental frames at two void fractions are shown in Fig.~\ref{fig:data_frames}.

Bubble images were recorded using a Photron FASTCAM Nova R2 camera equipped with a Navitar Resolv4K high-magnification lens and synchronized with a uniform tungsten--halogen backlight  (OSL2, Thorlabs). The imaging system was operated at  exposure time of 2.78~$\mu\mathrm{s}$, providing a field of view of approximately \(6\times6~\mathrm{mm}^{2}\) and a spatial resolution of \(2.9\pm0.1~\mu\mathrm{m\,pixel^{-1}}\). Images were stored in a lossless format so that the grayscale intensity, edge sharpness, halo structure, and other defocus-dependent optical signatures were retained for comparison with the ray-traced images. Spatial calibration was performed using a reference wire of diameter \(250.0\pm0.5~\mu\mathrm{m}\), yielding an estimated positional accuracy of approximately \(6~\mu\mathrm{m}\) for sharply resolved interfaces.

For the original experimental characterization, the raw frames were background-subtracted and contrast-enhanced before bubble contours were identified using Canny edge detection and overlapping bubbles were separated using watershed segmentation~\citep{kumar2026bubble}. The projected area \(A_i\) of each detected bubble was converted to an area-equivalent diameter, $d_i=(4A_i/\pi)^{1/2}$. More than \(20000\) bubbles (until statistics converged) were analyzed for each operating condition to establish statistically converged bubble-size distributions and characteristic diameters. In the present study, however, these previously processed measurements are used only as independent population-level references. The inverse model is applied directly to the experimental grayscale images using the preprocessing and instance-extraction procedure described below; experimentally inferred bubble depths or diameters are not supplied as bubble-level training labels. A preliminary depth-aware analysis of this experimental dataset was reported in Ref.~\citep{nayak2026depth}, whereas the present work introduces DNS-resolved supervision, experimentally validated ray tracing, joint physical-depth and diameter reconstruction, and closed-loop three-dimensional validation.

\subsection{DNS-resolved three-dimensional bubble dataset}
\label{sec:dns_dataset}

The three-dimensional physical data used to generate the synthetic shadowgraphs were obtained from direct numerical simulations of dilute bubbly decaying homogeneous isotropic turbulence. The simulations resolve the coupled evolution of the carrier-phase turbulence and the dispersed interfaces, including bubble deformation, relative motion, collision, coalescence, and breakup, without introducing wall effects or mean shear. The incompressible two-phase equations were solved using the accurate conservative diffuse-interface (ACDI) phase-field formulation, which maintains a bounded phase indicator and preserves the gas--liquid interface during large deformation and topology change~\citep{jain2020,jain2022a}. The governing equations were discretized on a staggered Cartesian grid using a low-dissipation, second-order skew-symmetric finite-volume formulation. Time advancement employed a fourth-order Runge--Kutta scheme, and incompressibility was enforced using a fractional-step pressure projection. All simulations were performed using the GPU-accelerated \textsc{ExaFlow3D} solver~~\citep{jain2020,jain2022a,jain2025stationary}. Complete details of the governing equations, numerical formulation, grid-convergence assessment, and bubble-population dynamics are provided in the companion study~\citep{kumar2026scaling,kumar2026cascade}.

The simulations were conducted in a triply periodic cubic domain representing freely decaying bubbly turbulence. The initial velocity field was generated from a prescribed three-dimensional energy spectrum whose large-scale turbulence statistics were calibrated against measurements in the pump-driven turbulent duct flow described above. This procedure reproduced experimentally relevant large-scale turbulence without directly imposing the measured velocity field. At the time at which the dispersed phase was introduced, the carrier turbulence had a Taylor-scale Reynolds number of
\(\mathrm{Re}_{\lambda}=206\), turbulent kinetic energy
\(k=1.19~\mathrm{m^{2}\,s^{-2}}\), and dissipation rate
\(\varepsilon=225~\mathrm{m^{2}\,s^{-3}}\). These values correspond to an integral length scale of approximately \(3.1~\mathrm{mm}\) and an initial Kolmogorov scale of approximately \(8~\mu\mathrm{m}\). The physical domain size was
\(L_{\mathrm{box}}\simeq15.6~\mathrm{mm}\), approximately five integral length scales, and the production simulations employed a uniform \(1024^{3}\) grid. The interface thickness was prescribed as \(0.51\Delta\), where \(\Delta\) is the grid spacing. The minimum value of \(\kappa_{\max}\eta\) for the finest grid was approximately \(1.65\), and it increased as the turbulence decayed, demonstrating that the dynamically relevant dissipative scales remained resolved over the analysis interval. For two-phase resolution, we have verified that $K_{\max} \times d_H \ge 60$~\cite{hatashita2026scalings}. The integral-scale Weber number was \(\mathrm{We}_{\ell}=63\), and the density and viscosity ratios were selected to represent an air--water system~\citep{kumar2026bubbles}. We used two direct numerical simulations (DNS) in diluted  void fractions regime,
\(\langle\phi\rangle=0.5\%\) and \(2\%\), were simulated. For each case, the initial bubble number and diameter distribution were prescribed using the statistically converged experimental bubble-size distributions. Bubbles were initially placed randomly throughout the periodic domain, subject to the constraint that no two interfaces overlapped or were in contact. They were introduced in nearly spherical form so that their subsequent deformation and topology changes arose dynamically from interaction with the resolved turbulent field.

The initial bubble population was not restricted to capillary-stable, sub-Hinze bubbles. The size distribution was deliberately selected to span the instantaneous Hinze scale,
\begin{equation}
    d_H =
    \mathrm{We}_{c}^{3/5}
    \left(\frac{\gamma}{\rho_{\ell}}\right)^{3/5}
    \varepsilon^{-2/5},
\end{equation}
\noindent where \(\gamma\) is the surface tension, \(\rho_{\ell}\) is the liquid density, and \(\mathrm{We}_{c}\) is the critical Weber number~\citep{hinze1955fundamentals,martinez1999breakup}. Initially, \(d_{32}\) (Sauter mean diameter) was close to \(d_H\), while the upper tail and the largest bubbles extended above \(d_H\). The DNS therefore began with both sub-Hinze and super-Hinze populations, representative of a bubble distribution produced by intense turbulent fragmentation. The higher-void-fraction case contained a broader upper tail and retained a larger super-Hinze population.

Following introduction into the turbulent field, the initially spherical bubbles underwent rapid strain-induced deformation. In particular, super-Hinze bubbles were strongly stretched before either fragmenting or merging with neighboring bubbles. Their transient geometries included elongated, flattened, irregular, and partially necked interfaces, together with shapes produced immediately before and after collision, breakup, and coalescence. These deformations modify the instantaneous projected area, interface curvature, collision cross-section, and optical signature and are not represented adequately by spherical synthetic particles. The DNS states selected for ray tracing were therefore deliberately drawn from multiple stages of the bubble-population evolution and included a substantial number of strongly deformed bubbles. The selected database spans the early mixed breakup--coalescence interval, during which a finite super-Hinze population remained active, as well as the later predominantly sub-Hinze, coalescence-dominated interval.

Individual bubbles were identified at every selected DNS timestamp using connected components of the resolved gas-phase indicator field. For bubble \(i\), its volume was evaluated directly from the phase field, and its volume-equivalent physical diameter was defined as
\begin{equation}
V_{i}=\int_{\Omega_i}\phi_g(\mathbf{x})\,\mathrm{d}V,
\qquad
d_i=\left(\frac{6V_i}{\pi}\right)^{1/3},
\end{equation}
\noindent where $\phi_g$ is gas fraction and $\Omega_i$ is total volume of simulation.
This definition preserves the resolved gas volume and provides an unambiguous scalar size label even when the instantaneous interface is strongly non-spherical. This volume also accounts for the correctness due to the diffuse and finite interface thicknesses~\citep{nathan2025accurate}. The bubble-center coordinates were obtained from the corresponding volume-weighted first moments. The volume-equivalent diameter was used only as the physical regression target; the actual DNS-resolved interface geometry, including deformation and local surface structure, was retained when constructing the three-dimensional scene supplied to the forward ray tracer. The synthetic images were therefore generated from the physically evolved bubble shapes.

For each selected DNS time step, the bubble-center coordinates, resolved interface geometry, volume-equivalent diameters, timestamp, and bubble identities were retained. The same three-dimensional state was subsequently rendered from multiple camera orientations and focal-plane positions. This construction provided exact bubble-level correspondence between the physical DNS field and every ray-traced optical observation while exposing the inverse model to coupled variations in diameter, shape, depth, defocus, orientation, and projected overlap. The DNS labels were used only for supervised training and quantitative validation; they were not supplied to the network during inference.

\section{Ray-tracer framework}
\label{sec:forward_framework}

The synthetic shadowgraphs used to build the labeled dataset of Sect.~\ref{sec:dns_dataset} are generated by a Monte Carlo ray-transport simulator that renders a DNS-resolved bubble scene through a physically explicit model of the imaging system used in the experimental facility. The inverse model of Sect.~\ref{sec:inverse_framework} is trained entirely on these renders and applied to experimental images without further tuning (Sect.~\ref{subsec:experimental_transfer}), so the fidelity of the forward model conditions the transfer results reported in Sect.~\ref{sec:validation_methodology}. The approach follows the strategy demonstrated recently for transparent refractive objects of complex shape \citep{kalita2026_ray_tracing_transparent}. This section states the model, the optical cues available to the inverse network, and the validation results. All parameters required to reproduce the renders are collected in Table~\ref{tab:raytracer_config}.

\begin{figure}[h]
    \centering
    \includegraphics[width=0.7\columnwidth]{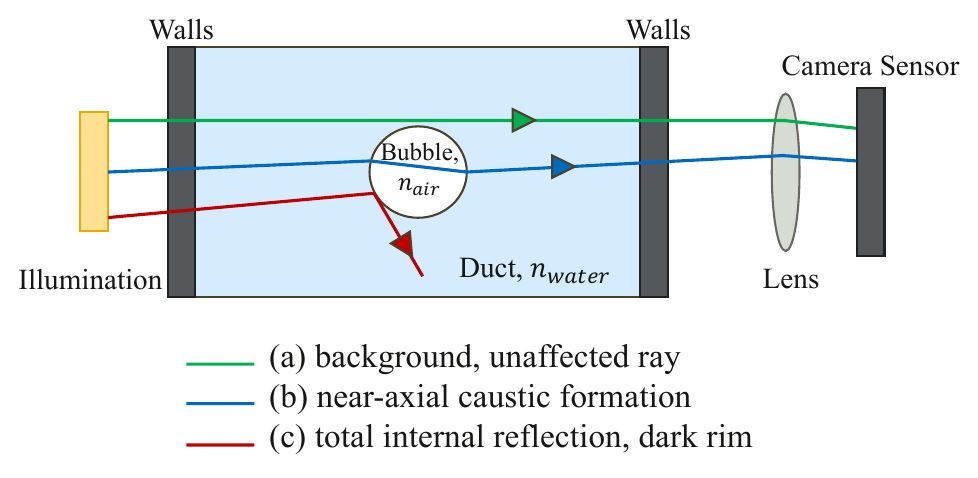}
    \caption{
    Optical stack and representative ray paths, not drawn to scale. Light propagates from the illumination source through the water channel to the lens and sensor. Ray (a) is unaffected by any bubble. Ray (b) passes near the bubble axis with little deviation and reaches the lens, forming the near-axial caustic. Ray (c) strikes the bubble beyond the normalized impact parameter $\sin\theta_i=0.750$ and undergoes total internal reflection on the bubble's outer surface (Eq.~\ref{eq:tir_angle}), contributing to the dark rim.
    }
    \label{fig:optical_schematic}
\end{figure}

\subsection{Bubble optics and diffraction regime}
\label{subsec:bubble_optics_regime}

An air bubble in water has relative refractive index
\begin{equation}
m = \frac{n_\text{air}}{n_\text{water}} = \frac{1.000}{1.333} = 0.750,
\label{eq:relative_index}
\end{equation}
so the bubble acts as a diverging optical element. Total internal reflection occurs on the outer surface of the sphere, where light passes from water into air, at
\begin{equation}
\theta_c = \arcsin m = 48.61^\circ,
\label{eq:tir_angle}
\end{equation}
so rays striking the sphere beyond a normalized impact parameter $\sin\theta_i>0.750$, the outer $25\%$ of the projected radius and $44\%$ of the projected area, are specularly reflected. Inside the bubble the ray travels from the lower to the higher index, so total internal reflection cannot occur on exit. The transmitted image is therefore a dark disc. Droplets and glass beads, for which $m>1$, produce a bright-cored disc, and the depth-from-defocus signatures developed for those objects \citep{konig1986_glarepoint,vandehulstwang1991_glarepoints,glover1995_interferometric} apply to a bubble only after this inversion of contrast is accounted for.

\begin{figure*}[!t]
    \centering
    \includegraphics[width=\columnwidth]{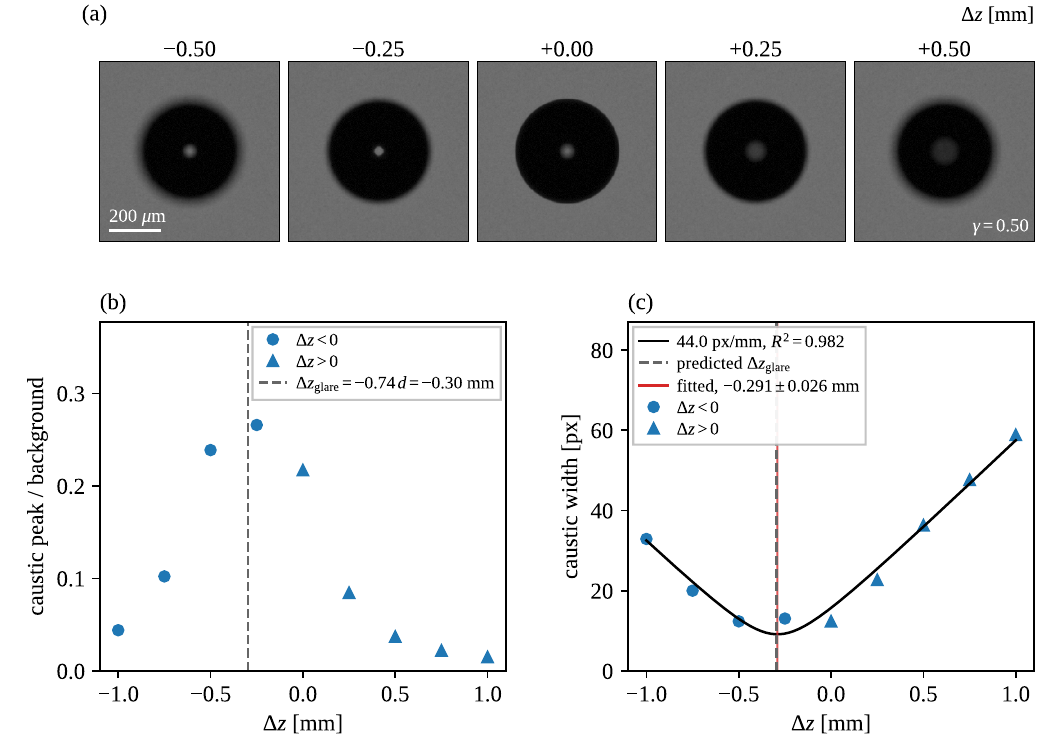}
    \caption{
    Optical signature of a single $0.40$~mm bubble across a signed
    focal-plane-relative depth sweep, on the traverse grid of
    Sect.~\ref{subsec:wire_validation}. All panels derive from the same nine
    renders. (a) Five of the nine depths on a common spatial and intensity
    scale, panel titles giving $\Delta z$ in millimetres. A display gamma of
    $0.50$ is applied to the display mapping and is stamped on the figure;
    the rendered data are unmodified. (b) Peak caustic transmittance
    normalized by the background. (c) Full width at half maximum of the same
    caustic, with the quadrature fit $w=[k^2(\Delta z-z_0)^2+w_0^2]^{1/2}$
    taken over all nine depths. Dashed grey line, the bubble position
    $\Delta z_\text{glare}$ at which Eq.~\eqref{eq:glare_offset} predicts the
    caustic to be in focus; solid red line, the fitted vertex $z_0$. Circles
    denote $\Delta z<0$ (near-lens side) and triangles $\Delta z>0$
    throughout. Measured values are given in
    Sect.~\ref{subsec:bubble_optics_regime}.
    }
    \label{fig:bubble_focus_sweep}
\end{figure*}

The bubble forms a virtual image of the backlight, observed as the near-axial caustic at the centre of the disc. Measured from the sphere centre, the paraxial focal length of a sphere of diameter $d$ is $f_b=md/[4(m-1)]=-0.75\,d$. The backlight lies at a standoff of $L=18.5$~mm behind the channel mid-plane (Table~\ref{tab:raytracer_config}), so the thin-lens relation places the virtual image a distance $0.75\,d\,L/(L+0.75\,d)$ behind the bubble centre, on the source side. The camera records this image in focus when it coincides with the focal plane, which occurs when the bubble centre lies at
\begin{equation}
\Delta z_\text{glare} = -\frac{0.75\,d\,L}{L+0.75\,d} \simeq -0.74\,d,
\label{eq:glare_offset}
\end{equation}
where $\Delta z=z_\text{bubble}-z_\text{focus}$ and $z$ increases from the lens toward the source. The offset is $-0.15$~mm for a $0.20$~mm bubble and $-0.51$~mm for a $0.70$~mm bubble, so the caustic is sharpest with the bubble on the near-lens side of the focal plane. Figure~\ref{fig:bubble_focus_sweep} tests this prediction on a $0.40$~mm bubble rendered at nine depths spanning $|\Delta z|\le1.00$~mm. The image strip of Fig.~\ref{fig:bubble_focus_sweep}(a) shows the caustic compact on the near-lens side of focus and diffuse at the matched positive depth. Panel~(b) gives the brightness. Peak transmittance reaches its maximum at $\Delta z=-0.25$~mm, displaced from focus, and at matched $|\Delta z|$ the near-lens side is brighter by factors of $3.1$, $6.4$, $4.7$ and $2.9$ at $0.25$, $0.50$, $0.75$ and $1.00$~mm. Panel~(c) gives the width. A fit of $w=[k^2(\Delta z-z_0)^2+w_0^2]^{1/2}$ over all nine depths, the same quadrature form applied to the calibration wire in Sect.~\ref{subsec:forward_validation_results}, returns $k=44.0\pm1.6$~px/mm, an in-focus floor of $w_0=9.1\pm1.9$~px, and a vertex at $z_0=-0.291\pm0.026$~mm, with a root-mean-square residual of $2.1$~px and $R^2=0.982$. The fitted vertex agrees with the $-0.295$~mm given by Eq.~\eqref{eq:glare_offset} within its uncertainty. The caustic therefore behaves as an object displaced by $0.74\,d$ toward the source and is in focus when the bubble sits at $\Delta z_\text{glare}$. It is a dim feature in absolute terms: its peak transmittance is $27\%$ of the unobstructed background at its own focus, so it appears as a small grey disc within a black shadow.

Over the $|\Delta z|\le1.00$~mm sweep the circle of confusion is symmetric about focus to within $1.4\%$ (rising to $9.9\%$ at the channel walls), so it constrains mainly the magnitude of the defocus. The sign is carried by the caustic, whose in-focus position scales with $d$ through Eq.~\eqref{eq:glare_offset}. Sign and diameter are therefore coupled, and a joint estimation model is required for both variables.

Light transport is treated geometrically, with diffraction retained as a single incoherent point-spread function. Four bounds evaluated for the present configuration support this treatment. The Mie size parameter $x=\pi d/\bar\lambda$ is $916$--$3206$ over the $0.20$--$0.70$~mm diameter range, so away from caustic directions the geometric limit of Lorenz--Mie scattering carries a relative error of order $x^{-2/3}$, at most about $1\%$ \citep{vandehulst1957_lightscattering}. The source is a tungsten--halogen bundle with roughly $300$~nm of effective bandwidth, giving a temporal coherence length $\bar\lambda^2/\Delta\lambda\approx1.6~\mu$m, so intensities, not fields, are summed over wavelength. By the van Cittert--Zernike theorem the transverse coherence radius of the $3.2$~mm-radius source at the bubble plane is $0.61\bar\lambda L/R_s\approx2.4~\mu$m, against a Fresnel-zone scale $\sqrt{\bar\lambda z}$ of $70~\mu$m for a bubble at the mid-plane and up to $100~\mu$m at the rear wall, $z$ being the propagation distance to the front wall, so edge-diffraction fringes are averaged out over the source and the shadowgraph description applies. Finally, the defocus wavefront error exceeds one wave beyond $|\Delta z|\approx0.33$~mm, and the geometric circle-of-confusion radius reaches $179$~px at the near-lens wall, while the Airy first-zero radius varies only between $2.0$ and $3.1$~px over the channel depth. A single point-spread function evaluated at the conjugate plane therefore serves the whole frame.

\subsection{Forward model}
\label{subsec:forward_model_summary}

The optical axis $z$ runs from the lens ($z=0$) toward the source through five plane-parallel layers (Fig.~\ref{fig:optical_schematic}): a $100.0$~mm air gap, a $9.50$~mm polycarbonate window ($n=1.586$), the $13.97$~mm water channel ($n_3=1.333$), a second $9.50$~mm window, and a $2.0$~mm air gap to the face of the fiber bundle (Table~\ref{tab:raytracer_config}). The source is a uniform disc of radius $R_s=3.2$~mm with fiber numerical aperture $0.39$. Each DNS bubble enters the scene as a sphere of its volume-equivalent diameter at its DNS centre, translated so that the DNS domain centre coincides with the optical axis at the channel mid-plane. Bubbles whose centre lies outside the water layer or outside the field of view are discarded.

The simulator estimates a dimensionless transmittance. Two frames are rendered for every scene, one with the bubble population present and one with the channel empty (the calibration frame). The transmittance at pixel $j$ is
\begin{equation}
\mathcal T_j=\frac{\Phi_j^{(s)}}{\left(G_\sigma*\Phi^{(c)}\right)_j},
\label{eq:transmittance}
\end{equation}
where $\Phi_j^{(s)}$ and $\Phi_j^{(c)}$ are the weighted ray counts in the scene and calibration frames, and $G_\sigma$ is a Gaussian of $\sigma=1.5$~mm on the sensor that suppresses the Monte Carlo noise of the calibration frame while preserving the millimetre-scale illumination profile. Source radiance, exposure and the flat-interface Fresnel losses are common to both frames and cancel in expectation; the absolute scale is reintroduced through the background electron count of the sensor model below.

Rays are generated by two complementary samplers. A background sampler draws source points uniformly over the emitter disc and directions uniformly within a cone of half-angle $\theta_p=\arcsin\mathrm{NA}=4.30^\circ$ about the lens axis. A targeted sampler assigns $75\%$ of the ray budget, divided evenly among the bubbles, to a cap about each bubble $b$, with axis $\hat a_b=(\mathbf C_b-\mathbf x_s)/\|\mathbf C_b-\mathbf x_s\|$ recomputed for every sampled source point $\mathbf x_s$ and half-angle $\alpha_b=\arcsin[\kappa R_b/\|\mathbf C_b-\mathbf x_s\|]$ with margin $\kappa=2.5$. The targeted sampler is required because, for a bubble on the axis, rays from the edge of the source leave it at about $13.8^\circ$ in the rear air gap, against the $4.30^\circ$ background cone, so most of the flux that a bubble delivers to the pupil leaves the source outside that cone. With $N_u$ background rays, $N_c$ rays per cap, $N=N_u+\sum_b N_c$, and cone and cap solid angles $\Omega_p$ and $\Omega_b$, the two samplers are combined by the multiple-importance-sampling balance heuristic \citep{veachguibas1995_mis},
\begin{equation}
w(\hat\omega)=\frac{N/\Omega_p}{\dfrac{N_u}{\Omega_p}\mathbb 1[\hat\omega\in\Omega_p]+\displaystyle\sum_b\frac{N_c}{\Omega_b}\mathbb 1[\hat\omega\in\Omega_b]},
\label{eq:mis_weight}
\end{equation}
which reduces to $w\equiv1$ for an empty scene, so that Eq.~\eqref{eq:transmittance} has unit expectation away from bubbles.

At each flat interface a ray is refracted by Snell's law and its weight is multiplied by the unpolarized Fresnel transmittance $T=1-\tfrac12(R_s+R_p)$; the reflected fraction is not traced. The four flat interfaces between source and lens (air/polycarbonate, polycarbonate/water, water/polycarbonate, polycarbonate/air) together transmit $88.6\%$ at normal incidence. At each bubble the ray is refracted on entry and on exit with the same Fresnel weighting. Rays incident beyond $\theta_c$ are reflected specularly with unit weight and traced onward. Internal reflections inside the bubble are not traced: the internal reflectance is about $2.0\%$ at normal incidence, and the first internal order that leaves toward the lens requires two internal reflections, carrying about $4\times10^{-4}$ of the transmitted flux. Bubbles are tested in a single pass in order of decreasing rear-surface depth, so a ray deflected by one bubble can still interact with bubbles that lie closer to the lens.

The lens is modeled as an equivalent paraxial system fixed by two calibrated numbers, the magnification $M$ and the object-side numerical aperture $\mathrm{NA}$ at the conjugate plane $u_0$, from which the focal length, entrance pupil diameter, and image distance follow as $f=Mu_0/(M+1)$, $D=2\,\mathrm{NA}\,u_0$ and $v_s=Mu_0$. The conjugate relation acts on the reduced optical path $u(z)=\int_0^z\mathrm dz'/n(z')$. The front window and the water both shorten the path seen by the lens, so the channel mid-plane sits at a physical $116.49$~mm and a reduced $u_0=111.23$~mm. A ray is retained if it crosses the lens plane within the circular stop of diameter $D$, and its sensor coordinate follows from the thin-lens transfer
\begin{equation}
X_s=X_\ell+v_s\left(\theta_x-\frac{X_\ell}{f}\right)=-\frac{v_s}{u}\,h+v_s\left(\frac1u-\frac1{u_0}\right)X_\ell,
\label{eq:sensor_transfer}
\end{equation}
where $X_\ell$ is the ray height at the lens, $\theta_x$ its slope, and $h$ the height of the object point at reduced distance $u$. At $u=u_0$ the second term vanishes and every ray from the point lands at $-Mh$; away from it the spread of $X_\ell$ over the pupil produces the circle of confusion, so defocus arises from the ray transport itself.

The ray histogram is convolved with a single polychromatic Airy point-spread function, integrated over $400$--$950$~nm with weights $S(\lambda)Q(\lambda)$ given by a $3200$~K blackbody spectrum $S$ and the sensor quantum efficiency $Q$; the corresponding mean wavelength is $\bar\lambda=686$~nm. The result is scaled by the background electron count $N_e\approx9.7\times10^3~e^-$, set by calibrating the source irradiance against the experimental background level, corrupted by Poisson shot noise and $12~e^-$ Gaussian read noise, normalized by the $5\times10^4~e^-$ full well, clipped, and written as 8-bit values, the format in which the experimental images are stored.

\subsection{Depth and diameter transfer}
\label{subsec:depth_diameter_cues}

From Eq.~\eqref{eq:sensor_transfer}, the circle-of-confusion radius on the sensor, in pixels of pitch $\Delta$, is
\begin{equation}
r_\text{CoC}(u)=\frac{D\,v_s}{2\Delta}\left|\frac1u-\frac1{u_0}\right|,
\label{eq:coc_radius}
\end{equation}
which grows with physical depth $z$ inside the immersion medium of index $n_3$ at
\begin{equation}
\frac{\mathrm dr_\text{CoC}}{\mathrm dz} = \frac{M}{\Delta}\tan\!\left[\arcsin(\mathrm{NA}/n_3)\right] = 24.50~\text{px/mm}.
\label{eq:coc_growth_rate}
\end{equation}
Equation~\eqref{eq:coc_growth_rate} agrees within $0.2\%$ with the first-order derivative $M\,\mathrm{NA}/(n_3\Delta)=24.46$~px/mm of Eq.~\eqref{eq:coc_radius}. The slope is set by $M$, $\mathrm{NA}$ and $n_3$ alone and is independent of where the lens's internal pupil sits. Evaluated in air, $n_3=1$, the same expression gives $32.70$~px/mm at the same aperture; this is the geometric rate expected for the drained-channel wire test of Sect.~\ref{subsec:wire_validation}. The inverse network of Sect.~\ref{subsec:multitask_inverse_model} learns these defocus signatures from each crop.

The apparent magnification also varies with depth. The sensor is fixed at $v_s$, thus the magnification that scales an object's projection is the chief-ray value $M_\text{proj}(u)=v_s/u$ of Eq.~\eqref{eq:sensor_transfer}, whose logarithmic gradient is $-1/u_0=-0.899\%$ per millimetre of reduced path, equivalently $-0.674\%$ per millimetre of physical depth in water. This is the magnification that enters the diameter correction of Eq.~\eqref{eq:diameter_correction}. The alternative quantity $v(u)/u$ describes the magnification of the sharp image an object forms at its own conjugate plane and applies when the camera is translated until the object is brought into focus; its logarithmic depth gradient, $-1/(u_0-f)$, is larger by the factor $u_0/(u_0-f)=M+1=5.35$. The two definitions coincide at $u=u_0$, which fixes the definition of the calibrated $M$ of Table~\ref{tab:raytracer_config}. The perspective term changes sign with the defocus, increasing the apparent radius on one side of focus and decreasing it by the same amount on the other, while blur produces, to first order, the same increase at equal positive and negative defocus. The perspective term therefore carries the sign. Its magnitude is small: for a $0.40$~mm bubble it changes the apparent radius by $\pm1.35~\mu$m, or $\pm0.6$ object-space pixels of $2.3~\mu$m, over $\pm1$~mm of depth in water, against up to $56~\mu$m of blur growth. It is therefore treated as a secondary cue.

Once the network produces a depth estimate $u$, the recovered bubble diameter is corrected for the defocus contribution to the apparent size,
\begin{equation}
d_\text{true} = \frac{\Delta}{M_\text{proj}(u)}\left[d_\text{apparent}-2r_\text{CoC}(u)\right],
\label{eq:diameter_correction}
\end{equation}
in which $d_\text{apparent}$ is twice the outermost radius, in sensor pixels, at which the shadow recovers to $95\%$ of the local background, and $r_\text{CoC}(u)$ is given by Eq.~\eqref{eq:coc_radius}.

\begin{table}[!t]
\caption{Ray-tracer configuration. The values above the first rule are calibrated or instrumental; those between the rules follow from them through the equivalent-lens relations of Sect.~\ref{subsec:forward_model_summary}; those below the second rule define the stack, source, sampling and sensor models.}
\label{tab:raytracer_config}
\begin{tabular}{@{}lll@{}}
\toprule
Symbol & Quantity & Value \\
\midrule
$\Delta$ & pixel pitch & $10.0~\mu$m \\
 & sensor format & $2048\times2048$ \\
 & object-space pixel scale & $2.300~\mu$m/px \\
$M$ & magnification & $4.348$ \\
$\mathrm{NA}$ & object-side numerical aperture & $0.075$ \\
$\bar\lambda$ & mean wavelength of the PSF spectrum & $686$~nm \\
$t_\text{exp}$ & exposure & $2.78~\mu$s \\
$N$ & rays per frame & $2.5\times10^9$ \\
\midrule
$u_0$ & reduced conjugate distance (from the stack) & $111.23$~mm \\
$f$ & equivalent focal length & $90.43$~mm \\
$D$ & entrance pupil diameter & $16.68$~mm \\
$v_s$ & image distance & $483.63$~mm \\
\midrule
 & air gap, lens to front window & $100.0$~mm \\
 & window thickness (polycarbonate, $n=1.586$) & $9.50$~mm \\
 & channel depth (water, $n_3=1.333$) & $13.97$~mm \\
 & air gap, rear window to source & $2.0$~mm \\
$L$ & source standoff from channel mid-plane & $18.5$~mm \\
$R_s$ & source radius & $3.2$~mm \\
 & source fiber numerical aperture & $0.39$ \\
$\kappa$ & cap margin of the targeted sampler & $2.5$ \\
 & fraction of rays in targeted caps & $0.75$ \\
$\sigma$ & calibration-frame smoothing & $1.5$~mm \\
$N_e$ & background electron count (flow stack) & $9.7\times10^3~e^-$ \\
 & read noise / full well & $12~e^-$ / $5\times10^4~e^-$ \\
\botrule
\end{tabular}
\end{table}

A single measurement convention is applied throughout this study, including the calibration-wire analysis. The averaged intensity profile across the edge (azimuthally averaged for a bubble, row-averaged for the straight wire) is differentiated to form a line spread function, a Gaussian is fitted to that function, and the blur radius is defined as half of the fitted full width at half maximum. The convention returns values below the geometric blur radius, since the defocus kernel of a uniform disc follows a chord-length distribution while the fitted model is Gaussian. The resulting bias is measured and reported as the transfer factor $\beta$ of Eq.~\eqref{eq:beta_wire}. The $50\%$-transmission crossing is used only to measure magnification on the straight wire, where it is unbiased under any even kernel narrower than the wire.

\begin{figure*}[t]
    \centering
    \includegraphics[width=\columnwidth]{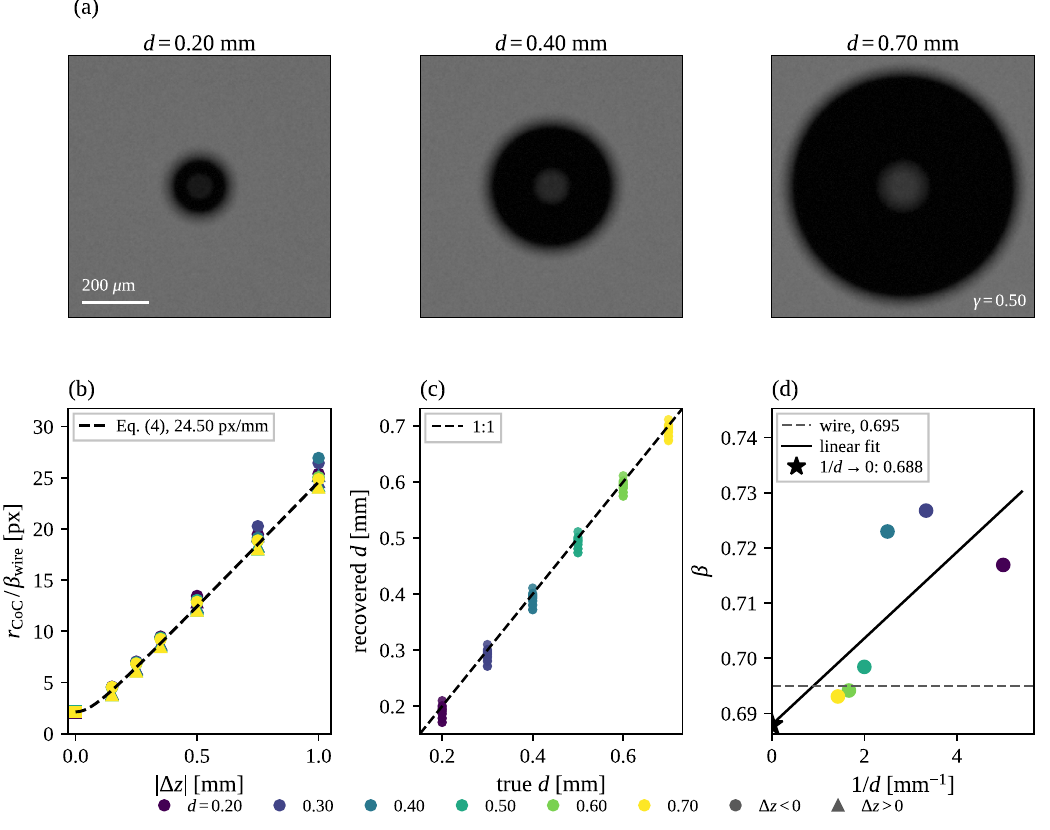}
    \caption{
    Self-consistency of the depth and diameter cues, evaluated entirely in
    simulation over six diameters, $d=0.20$--$0.70$~mm, and thirteen signed
    depths spanning $|\Delta z|\le1.00$~mm. All $78$ rendered points are
    plotted and fitted. (a) Rendered bubbles at three diameters, all at
    $\Delta z=+0.50$~mm, on a common spatial and intensity scale and shown
    with the same display gamma of $0.50$ as Fig.~\ref{fig:bubble_focus_sweep};
    the stretch is applied to the display mapping and is stamped on the
    figure, so the measurements in the remaining panels are drawn from the
    unmodified renders. (b) Blur radius, measured as half the full width at
    half maximum of a Gaussian fitted to the radial line spread function and
    divided by the transfer factor $\beta_\text{wire}$ of
    Eq.~\eqref{eq:beta_wire}; the dashed curve is
    Eq.~\eqref{eq:coc_growth_rate} combined in quadrature with the measured
    in-focus floor. (c) Diameter recovered by
    Eq.~\eqref{eq:diameter_correction} against the known input, with
    $d_\text{apparent}$ taken as twice the outermost radius at which the shadow
    recovers to $95\%$ of the local background and $r_\text{CoC}$ from
    Eq.~\eqref{eq:coc_radius}; the dashed line is the identity. (d) Transfer factor
    $\beta$ against inverse diameter $1/d$, which is proportional to edge
    curvature, with the least-squares line through all six diameters, the
    independently measured wire value dashed, and the extrapolation to
    $1/d\to0$ starred. Circles denote
    $\Delta z<0$ and triangles $\Delta z>0$. Measured values are given in
    Sect.~\ref{subsec:depth_diameter_cues}.
    }
    \label{fig:depth_diameter_cues}
\end{figure*}

Equation~\eqref{eq:coc_growth_rate} predicts a slope set solely by the imaging geometry, and Eq.~\eqref{eq:diameter_correction} inverts that relation to recover a size, so both are evaluated for internal self-consistency against the renderer's pixel output. This assessment is computational, since no isolated single-bubble experimental image is available for direct comparison (Sect.~\ref{subsec:validation_strategy}). It complements the two experimental validations available, the calibration-wire test of Sect.~\ref{subsec:wire_validation} and the closed-loop DNS reconstruction test of Sect.~\ref{subsec:closed_loop_validation}. Figure~\ref{fig:depth_diameter_cues} implements the check by rendering isolated single bubbles over a systematic diameter--depth grid and measuring the cues from the rendered pixels.

Figure~\ref{fig:depth_diameter_cues}(a) makes the diameter-independence of Eq.~\eqref{eq:coc_growth_rate} visible: across three bubbles rendered at a common depth of $\Delta z=+0.50$~mm the projected disc grows by a factor of $3.5$ while the width of the shadow edge changes by $6\%$. Panel~(b) quantifies it. A fit of $r_\text{CoC}=(k^2\Delta z^2+r_0^2)^{1/2}$ to each diameter separately, divided by the wire transfer factor of Eq.~\eqref{eq:beta_wire}, returns $k/\beta_\text{wire}=24.99$~px/mm on average, $2.0\%$ above the analytic value, within $4.6\%$ at every diameter and with a spread of $2.0\%$ across the diameter range; the measured in-focus floor is $1.43\pm0.06$~px. Panel~(c) applies Eq.~\eqref{eq:diameter_correction} to the same renders and recovers the input diameter with a root-mean-square error of $0.0147$~mm and a bias of $-0.0098$~mm over all $78$ points, $3.3\%$ of the mean diameter and $7.3\%$ of the smallest.

Panel~(d) tests whether the wire transfer factor applies to curved bubble edges. For each diameter, $\beta$ is the fitted slope $k$ of panel~(b) divided by the analytic rate of Eq.~\eqref{eq:coc_growth_rate}. The Gaussian convention responds to the shape of the edge profile, and a curved edge spreads blurred light unequally to its inner and outer sides, so $\beta$ depends on the edge curvature, which scales as $1/d$. Across the six diameters it ranges from $0.693$ at $d=0.70$~mm to $0.727$ at $d=0.30$~mm. A least-squares line in $1/d$ through all six diameters extrapolates at $1/d\to0$, the straight-edge limit, to $\beta=0.688$, within $1.0\%$ of the value measured independently on the straight wire (Eq.~\eqref{eq:beta_wire}). The correction applied in panel~(b) is therefore a property of the width estimator that transfers between the two geometries.

\subsection{Computational implementation}
\label{subsec:computational_implementation}

The production configuration renders $N=2.5\times10^9$ rays per frame across the complete labeled dataset of Sect.~\ref{sec:dns_dataset} ($900$ frames spanning the selected DNS timestamps, camera views, and focal-plane settings) on GPU-accelerated hardware (NVIDIA H200). Rays are generated on the device and traced in parallel, one GPU thread per ray stream, with atomic accumulation into the sensor histogram; the calibration smoothing, point-spread function and sensor noise are applied afterwards on the host. The simulator is implemented in three backends that share the same physics and differ only in their random-number streams: a vectorized NumPy CPU path, a CuPy array-drop-in GPU path, and a fused Numba CUDA-kernel GPU path, the last of which produced all production frames. Frames are independent, so the dataset is generated as a job array with one frame per task and no inter-task communication.

\subsection{Validation strategy}
\label{subsec:validation_strategy}

The forward model is validated in two complementary ways. Isolation of a single bubble in the present turbulent flow lies outside the capability of the current experimental configuration, and both tests are constructed around this constraint. The imaging chain of Sect.~\ref{subsec:forward_model_summary}, comprising the paraxial mapping, the defocus blur and the sensor radiometry, is validated against an opaque calibration wire of exactly known geometry. The wire transmits no light, so the imaging physics is exercised in isolation from bubble refraction (Sect.~\ref{subsec:wire_validation}). The refractive ray transport of Sect.~\ref{subsec:forward_model_summary} requires a separate treatment, since the duct flow provides no experimentally accessible, well-characterized single refractive object of the kind available in the isolated levitated and free-falling drop experiments used to validate refractive ray tracing elsewhere \citep{kalita2026_ray_tracing_transparent}. This physics is validated through the closed-loop DNS-to-image-to-reconstruction test of Sect.~\ref{subsec:closed_loop_validation}. A DNS-resolved bubble field of known three-dimensional geometry is rendered by the same forward model and passed through the complete inverse-reconstruction cycle, so that any unmodeled refractive bias appears as bubble-level depth or diameter error measured against ground truth.

The wire renders use the background sampler alone. Of the $2.5\times10^9$ rays, $47\%$ reach the sensor, so a pixel near the frame centre accumulates about $4\times10^2$ rays and carries a Monte Carlo relative error of about $5\%$, against a photon shot-noise floor of $1/\sqrt{N_e}\approx0.75\%$ for the brighter drained stack, $N_e\approx1.8\times10^4~e^-$. In the raw histogram this term is independent between pixels, since each ray deposits its weight into a single pixel. Averaging over the $N_r=400$ rows of the edge profile of Sect.~\ref{subsec:wire_validation} reduces it to about $0.25\%$ of the background level. The measured simulation-to-experiment root-mean-square deviation is $3.39\%$, so the sampling term accounts for about $0.5\%$ of that residual in variance. Averaging reduces random variations but does not reduce systematic errors in the forward model. Consequently, the residual observed in the wire comparison is primarily due to model error. Individual bubbles, in the current configuration, span only a few tens of pixels, so averaging is less effective for these features. As a result, sampling effects make a proportionally larger contribution to the closed-loop test described in Sect.~\ref{subsec:closed_loop_validation}.

\subsection{Calibration-wire validation}
\label{subsec:wire_validation}

The imaging model is validated using a $500$-$\mu$m opaque wire imaged with the channel drained. In the experiment the wire sat $111.25$~mm from the lens at sharp focus, the conjugate distance of the channel mid-plane; the wire renderer places it at its own conjugate distance with the same magnification and numerical aperture, so the comparison is made at identical $M$ and $\mathrm{NA}$. The wire is surrounded by air, so its geometric blur grows at the air-stack rate of $32.70$~px/mm. Defocus is produced by translating the camera (lens and sensor) along the axis relative to the wire, windows and backlight, so the illumination geometry seen by the wire is identical at every setting. Nine defocus settings were used, $\Delta z\in\{0,\pm0.25,\pm0.50,\pm0.75,\pm1.00\}$~mm; the eight defocused settings enter the origin-forced aperture fit, and all nine enter the three-parameter fit of Sect.~\ref{subsec:forward_validation_results}.

In the geometric regime the defocus kernel projected onto a line is the chord-length distribution of a uniform disc of radius $r_\text{CoC}$, a semi-ellipse of full width at half maximum $\sqrt3\,r_\text{CoC}$. The Gaussian convention of Sect.~\ref{subsec:depth_diameter_cues} therefore reports a width below the geometric value, and the size of the bias depends on how the Gaussian is fitted: matching second moments gives a transfer factor of $0.589$, a least-squares fit with amplitude fixed at the peak over the support of the semi-ellipse gives $0.888$, and the deployed free-amplitude fit over a window wider than the support gives $0.700$. The physical aperture is obtained by rescaling the input aperture by the ratio of the experimental and simulated estimates, which is valid because the transfer factor does not depend on the aperture. The construction requires that the bias be common to both datasets, which holds because the same extraction code is applied to each and the geometric term dominates the kernel for $|\Delta z|\ge0.50$~mm; the origin-forced fit, weighted by $|\Delta z|$, limits the influence of the $\pm0.25$~mm settings, where the instrument blur floor is comparable to the geometric term. Magnification is measured separately from the $50\%$-transmission crossings of the two wire edges.

\subsection{Forward-model validation results}
\label{subsec:forward_validation_results}

Simulated frames are rendered at the production ray budget $N=2.5\times10^9$ (Table~\ref{tab:raytracer_config}). Experimental frames are trial-averaged over repeated acquisitions at each defocus setting. Both datasets are processed by the same extraction routine.

\begin{figure*}[t]
    \centering
    \includegraphics[width=0.9\textwidth]{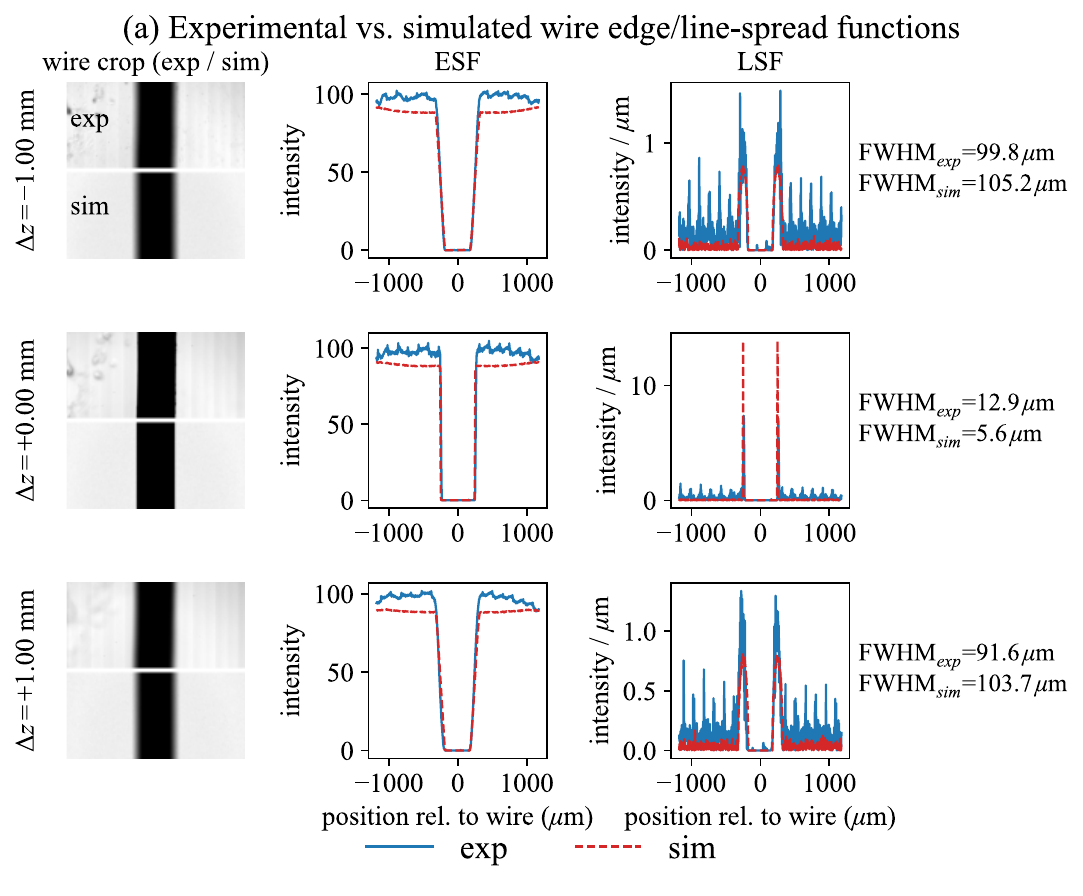}
    \caption{
    Calibration-wire images and edge response at three representative
    defocus settings, experiment against simulation. Rows correspond to
    $\Delta z=-1.00$, $0$ and $+1.00$~mm. Left column, the recorded and
    rendered crops, each display-normalized; middle column, the
    row-averaged edge spread function; right column, the corresponding line
    spread function, whose Gaussian full width at half maximum, halved, is
    the blur radius used throughout
    (Sect.~\ref{subsec:depth_diameter_cues}). The wire is $500~\mu$m in
    diameter and is imaged with the channel drained. Measured values are
    given in Sect.~\ref{subsec:forward_validation_results}.
    }
    \label{fig:wire_validation}
\end{figure*}

\begin{figure}[!t]
    \centering
    \includegraphics[width=0.6\columnwidth]{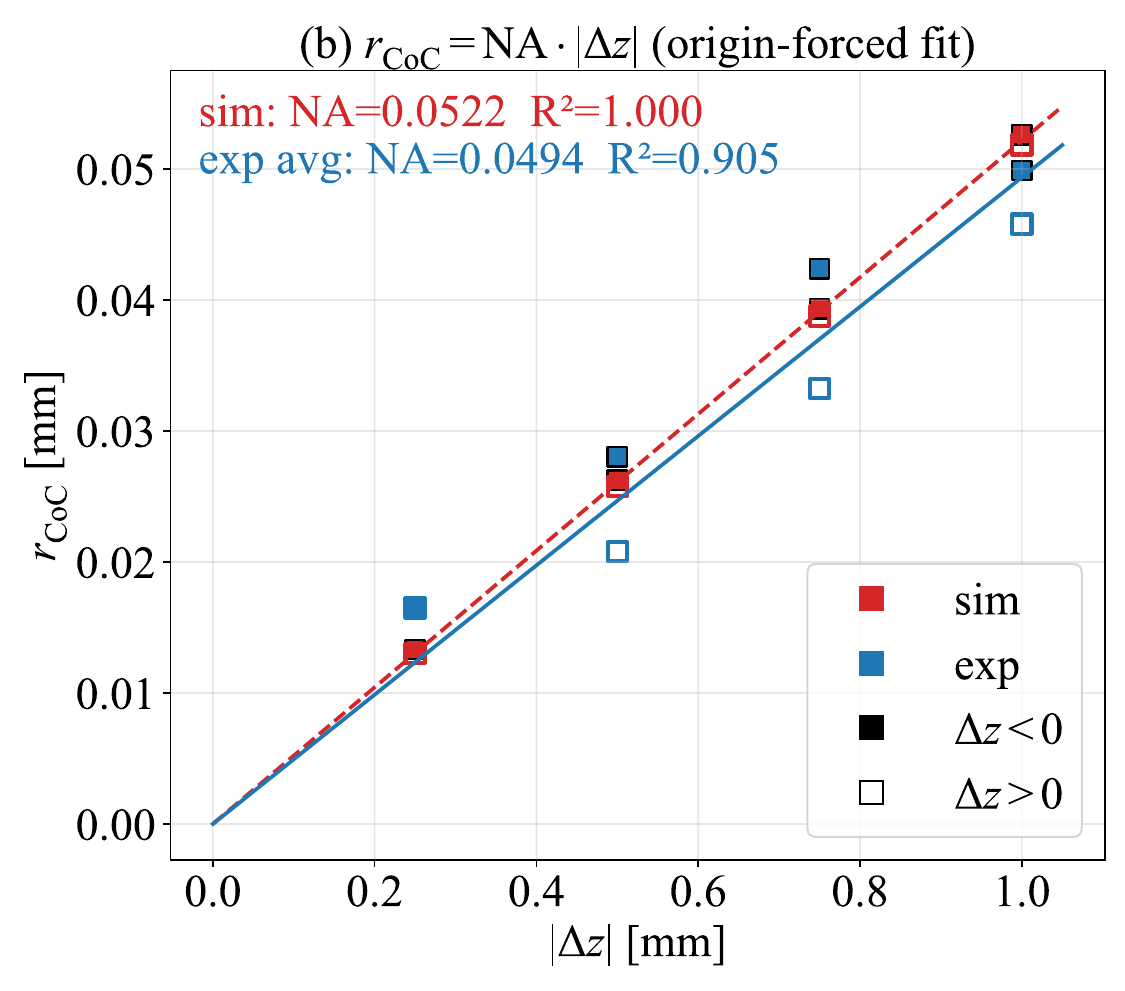}
    \caption{
    Blur radius against defocus magnitude for the eight defocused wire
    settings, experiment and simulation, with origin-forced linear fits.
    The channel is drained for this measurement, so the geometric slope at
    the rendered aperture is the air-stack value of $32.70$~px/mm, the
    in-water value of Eq.~\eqref{eq:coc_growth_rate} being $24.50$~px/mm;
    the plotted Gaussian radii grow at that slope multiplied by the transfer
    factor $\beta$. The ratio of the fitted slopes to the input aperture
    gives the transfer factor of Eq.~\eqref{eq:beta_wire}, and their ratio
    to each other gives the derived aperture reported in
    Sect.~\ref{subsec:forward_validation_results}.
    }
    \label{fig:wire_na_fit}
\end{figure}

Figure~\ref{fig:wire_validation} shows the paired experimental and simulated wire images with their edge- and line-spread functions at three representative settings. The shadow width at the $50\%$ transmission level, averaged over the nine settings, was $500.4~\mu$m measured and $500.1~\mu$m simulated against a certified wire diameter of $500~\mu$m. The mean per-frame deviation from the certified diameter is $0.24\%$ for the experiment and $0.30\%$ for the simulation, within the one-pixel resolution of the estimator, $\pm0.46\%$. The measured width therefore confirms $M=4.348$, equivalently an object-space scale of $2.300~\mu$m/px, to $0.1\%$ on average. After calibration of the source irradiance against the same images, the simulated background level averaged $90.5$ of $255$ written ADU against a measured $92.75$, a mean per-frame deviation of $2.86\%$ (maximum $5.87\%$). Normalized edge spread functions agreed with a mean Pearson correlation of $0.9964$ and a mean root-mean-square deviation of $3.39\%$ ($1.9$--$7.3\%$ per frame). Over the eight defocused settings the simulated Gaussian line-spread width differs from the measured one by $14.2\%$ on average ($24.0\%$ maximum), the largest differences occurring at $|\Delta z|\le0.50$~mm, where the instrument blur floor discussed below is comparable to the geometric term.

Figure~\ref{fig:wire_na_fit} plots the blur radius against defocus for the eight defocused settings, and its slopes yield the aperture. The origin-forced fit gives $\widehat{\mathrm{NA}}=0.0494$ from the experimental frames and $\widehat{\mathrm{NA}}=0.0522$ from the simulated frames at the input $\mathrm{NA}=0.075$, a relative difference of $5.69\%$. The realized transfer factor is therefore
\begin{equation}
\beta = \widehat{\mathrm{NA}}_\text{sim}/\mathrm{NA}_\text{in} = 0.695,
\label{eq:beta_wire}
\end{equation}
which agrees within $0.7\%$ with the $0.700$ that the deployed estimator returns for a purely geometric chord-length kernel. The diffraction and pixel terms of the simulated edge therefore perturb the estimator by under $1\%$ at these defocus settings, and the difference between $\beta$ and unity is located in the width convention. Equation~\eqref{eq:beta_wire} is the factor used in Fig.~\ref{fig:depth_diameter_cues}(b) and tested in Fig.~\ref{fig:depth_diameter_cues}(d). Rescaling the input aperture by $\widehat{\mathrm{NA}}_\text{exp}/\widehat{\mathrm{NA}}_\text{sim}$ returns a physical aperture of $0.0710$, $5.4\%$ below the value at which the sweep was rendered.

Two features of the experimental sweep bound how much of that residual is a blur-model error. The experimental origin-forced fit is poor, $R^2=0.905$ against $0.9996$ for the simulation, and the measured widths are asymmetric about the nominal focus, differing between the two signs by a factor of $1.35$ at $|\Delta z|=0.50$~mm where the simulation differs by $1.02$. A three-parameter fit admitting a focus offset and an in-focus blur floor, $\mathrm{FWHM}(\Delta z)=[k^2(\Delta z-\Delta z_0)^2+\mathrm{FWHM}_0^2]^{1/2}$, reduces both. It returns a traverse offset of $58\pm20~\mu$m in experiment against $5\pm1~\mu$m in simulation, which accounts for most of the observed asymmetry ($1.23$ of the measured $1.35$), with root-mean-square residuals of $4.04$ and $0.27~\mu$m respectively and an experimental $R^2$ of $0.980$. With that offset removed the measured and simulated blur slopes, $95.6\pm3.0$ and $104.0\pm0.2~\mu$m/mm, differ by $8.8\%$, a larger gap than the origin-forced $5.7\%$.

At the nominal focus, visible as the narrowest line spread functions in Fig.~\ref{fig:wire_validation}, the measured and simulated widths were $12.90$ and $5.56~\mu$m. The simulated value is accounted for to within $3\%$ by the $580$~nm Airy pattern used by the wire renderer convolved with the pixel aperture, so the simulation side is fully explained. The fitted in-focus floors of the three-parameter model, $16.5\pm4.0~\mu$m in experiment and $5.4\pm0.3~\mu$m in simulation, isolate $15.6\pm4.2~\mu$m of instrument blur absent from the model ($11.6~\mu$m from the nominal-focus frames alone). It is attributed to residual aberration of the physical zoom lens, which the equivalent paraxial model does not represent, and to sensor-level effects such as charge diffusion that lie outside the model. Expressed through the measured slope it is equivalent to $0.16\pm0.04$~mm of defocus and sets a floor on what the blur cue resolves near focus.

\section{Inverse bubble reconstruction framework}
\label{sec:inverse_framework}

The forward optical model described in the preceding section establishes the mapping from a DNS-resolved three-dimensional bubble field to its corresponding two-dimensional shadowgraphic observation. The inverse methodology addresses the complementary problem of recovering bubble position and physical size from the rendered or experimentally acquired image. Because the forward model preserves the identity, location, and physical-equivalent diameter of every simulated bubble, it provides paired optical and physical information for supervised development and controlled evaluation of the inverse reconstruction framework.

The premise that defocus itself carries recoverable depth information is well established in single-view optical sensing: image sharpness, blur radius, and the associated circle of confusion vary systematically with distance from the focal plane, and this dependence has long been exploited for depth-from-focus and depth-from-defocus estimation~\citep{pentland1987newsense}. Convolutional networks have since extended this principle to learned monocular depth prediction, in which a single image is mapped directly to a dense depth estimate without explicit multi-view geometry~\citep{eigen2014depthmap}. The present framework adapts this idea to a bubble-resolved regression setting: the network infers a single signed depth and a physical diameter for each detected bubble instance, with the known focal-plane position supplied as an auxiliary conditioning variable.

This inverse reconstruction framework also builds directly on a preliminary depth-aware pipeline developed for the same experimental facility, in which unsupervised Gaussian-mixture clustering of engineered focus-sensitive features (radial edge sharpness, gradient-weighted radial depth proxies, and projected scale) was combined with a supervised classifier to separate in-focus from out-of-focus bubbles~\citep{nayak2026depth,kartakoullis2025_gmm,xi2022_rf_dl}. That approach demonstrated that focus-dependent image cues carry usable depth information from a single camera. It produced a relative, unitless focus score and a binary in-focus/out-of-focus label, and its engineered features were tuned to the optical configuration in which they were derived. The present work replaces the engineered feature set and the unsupervised-clustering stage with DNS-resolved physical supervision, an experimentally validated ray-tracing forward model, and an end-to-end multitask convolutional network, and extends the earlier relative, qualitative focus ordering to a quantitative, physically calibrated, and experimentally transferable reconstruction of signed depth and physical-equivalent diameter.

Let \(I_{t,v,w}(u,v)\) denote a grayscale image associated with DNS timestamp \(t\), camera view \(v\), and focal-plane setting \(w\). The reconstructed bubble population is written as

\begin{equation}
\begin{aligned}
\widehat{\mathcal{B}}_{t,v,w}
&=
\left\{
\widehat{\mathbf{b}}_{i,t,v,w}
=
\left(
\widehat{x}_{i,t,v,w},
\widehat{y}_{i,t,v,w},
\widehat{z}_{i,t,v,w},
\right.\right.
\\
&\qquad\left.\left.
\widehat{d}_{i,t,v,w},
q_{i,t,v,w}
\right)
\right\}_{i=1}^{\widehat{N}_{t,v,w}},
\end{aligned}
\label{eq:inverse_population}
\end{equation}

\noindent where \(\widehat{N}_{t,v,w}\) is the number of detected bubble instances, \(\widehat{x}_{i,t,v,w}\) and \(\widehat{y}_{i,t,v,w}\) are the reconstructed lateral coordinates, \(\widehat{z}_{i,t,v,w}\) is the reconstructed physical depth, \(\widehat{d}_{i,t,v,w}\) is the reconstructed physical-equivalent diameter, and \(q_{i,t,v,w}\) is an instance-level confidence measure.

Recovering these quantities from a single image is intrinsically ill-posed because bubble appearance depends jointly on physical diameter and displacement from the focal plane. Different combinations of diameter and depth can produce similar apparent dimensions in the image plane. Defocus also alters edge sharpness, halo extent, boundary thickness, local contrast, and internal intensity structure. In dense bubbly flows, partial occlusion and overlap may obscure the visible boundary, and bubbles far from focus appear as diffuse shadows or low-contrast halos with no sharply closed contour. Apparent diameter alone is therefore insufficient to determine physical diameter or depth uniquely.

The inverse framework addresses this ambiguity through a hierarchical
procedure. Bubble-related image evidence is first localized and
separated into individual instances. The resulting segmentation
provides the spatial support required to preserve bubble identity,
define the bubble-centered crop, and map each prediction back to the
parent image. Quantitative depth and diameter inference is then
performed from the local grayscale optical signature together with
the known focal-plane coordinate. A multitask regression model jointly
estimates signed focal-plane-relative depth and physical-equivalent
diameter.

The true relative depth and the reconstructed absolute depth are defined by

\begin{equation}
\begin{aligned}
\Delta z_{i,t,v,w} &= z_{i,t} - z_f^{(w)}, \qquad \widehat{z}_{i,t,v,w} = z_f^{(w)} + \widehat{\Delta z}_{i,t,v,w},
\end{aligned}
\label{eq:relative_absolute_depth}
\end{equation}

\noindent where \(z_{i,t}\) is the physical depth of bubble \(i\), \(z_f^{(w)}\) is the known focal-plane depth, and \(\widehat{\Delta z}_{i,t,v,w}\) is the predicted relative depth.

This formulation separates the known imaging geometry from the residual depth information encoded in the bubble appearance. It also preserves the physical sign of defocus, allowing bubbles on opposite sides of the focal plane to be distinguished within a common reconstruction framework.

The complete inverse procedure consists of four coupled operations: localization of bubble-related image evidence, separation of individual bubble instances, joint prediction of relative depth and physical diameter, and assembly of the bubble-level predictions into a reconstructed three-dimensional population. Algorithm~\ref{alg:inverse_inference} summarizes this sequence at inference time, before the supervised construction and each individual stage are described in detail in the following subsections.

\begin{algorithm}[!htbp]
\caption{Inverse reconstruction inference for one image \(I_{t,v,w}\)}
\label{alg:inverse_inference}
\small
\begin{algorithmic}[1]

\Require grayscale image \(I_{t,v,w}\);
focal-plane depth \(z_f^{(w)}\);
trained operators
\(\mathcal{S}_{\psi}\),
\(\mathcal{I}_{\omega}\), and
\(\mathcal{G}_{\theta}\)

\Ensure reconstructed bubble population
\(\widehat{\mathcal{B}}_{t,v,w}\)

\State
\(P^{\mathrm{bub}}_{t,v,w}
\gets
\mathcal{S}_{\psi}(I_{t,v,w})\)
\Comment{localize bubble-related evidence}

\State
\(\{M^{\mathrm{inst}}_{i,t,v,w},q_{i,t,v,w}\}
\gets
\mathcal{I}_{\omega}
(I_{t,v,w},P^{\mathrm{bub}}_{t,v,w})\)
\Comment{separate bubble instances}

\State remove duplicate detections using confidence-based overlap suppression

\State
\(\widehat{\mathcal{B}}_{t,v,w}\gets\emptyset\)

\ForAll{retained instances \(i\)}

    \State determine visible and estimated physical support associated with
    \(M^{\mathrm{inst}}_{i,t,v,w}\)

    \State
    \(C_{i,t,v,w}
    \gets
    \mathcal{P}
    (I_{t,v,w},M^{\mathrm{inst}}_{i,t,v,w})\)
    \Comment{extract bubble-centered grayscale crop}

    \State
    \(\mathbf{r}_{i,t,v,w}
    \gets
    [\,\mathcal{N}_{C}(C_{i,t,v,w}),
    \widetilde{z}_{f}^{(w)}\,]\)

    \State
    \((\widehat{\Delta z}_{i,t,v,w},
    \widehat{d}_{i,t,v,w})
    \gets
    \mathcal{G}_{\theta}
    (\mathbf{r}_{i,t,v,w})\)

    \State
    \(\widehat{z}_{i,t,v,w}
    \gets
    z_f^{(w)}
    +
    \widehat{\Delta z}_{i,t,v,w}\)

    \State
    \((\widehat{x}_{i,t,v,w},
    \widehat{y}_{i,t,v,w})
    \gets
    \mathcal{T}
    (\widehat{u}_{i,t,v,w},
    \widehat{v}_{i,t,v,w},v)\)

    \State
    \(\widehat{\mathbf{b}}_{i,t,v,w}
    \gets
    (\widehat{x}_{i,t,v,w},
    \widehat{y}_{i,t,v,w},
    \widehat{z}_{i,t,v,w},
    \widehat{d}_{i,t,v,w},
    q_{i,t,v,w})\)

    \State
    \(\widehat{\mathcal{B}}_{t,v,w}
    \gets
    \widehat{\mathcal{B}}_{t,v,w}
    \cup
    \{\widehat{\mathbf{b}}_{i,t,v,w}\}\)

\EndFor

\State
\Return
\(\widehat{\mathcal{B}}_{t,v,w}\)

\end{algorithmic}
\end{algorithm}

\subsection{Synthetic supervision from DNS-resolved bubble fields}
\label{subsec:synthetic_supervision}

The supervised inverse problem is constructed from DNS-resolved bubble states and their corresponding ray-traced observations. At DNS timestamp \(t\), the physical bubble population is represented as

\begin{equation}
\mathcal{B}_{t}
=
\left\{
\mathbf{g}_{i,t}
=
\left[
x_{i,t},
y_{i,t},
z_{i,t},
d_{i,t}
\right]
\right\}_{i=1}^{N_t},
\label{eq:dns_population}
\end{equation}

\noindent where \(N_t\) is the number of bubbles, \((x_{i,t},y_{i,t},z_{i,t})\) are the bubble-center coordinates, and \(d_{i,t}\) is the physical-equivalent diameter.

For each selected DNS state, the forward optical model generates grayscale observations over multiple camera views and focal-plane settings. Each rendered frame is indexed by the combination \((t,v,w)\), where \(v\) denotes the camera view and \(w\) denotes the focal-plane setting with physical depth \(z_f^{(w)}\). This indexing is retained throughout segmentation, crop extraction, training, inference, and evaluation so that every bubble-level prediction remains linked to its source DNS state and optical configuration.

For bubble \(i\), the supervised regression target is

\begin{equation}
\mathbf{y}_{i,t,v,w}
=
\left[
\Delta z_{i,t,v,w},
d_{i,t}
\right]
=
\left[
z_{i,t}-z_f^{(w)},
d_{i,t}
\right],
\label{eq:training_target}
\end{equation}

\noindent where the first component is the signed focal-plane-relative depth defined in Eq.~\eqref{eq:relative_absolute_depth}, and the second component is the DNS-resolved physical-equivalent diameter. The target therefore preserves the known focal-plane geometry while requiring the model to infer the residual depth and physical size from the optical appearance.

Each crop is assigned a unique sample identity that records the DNS bubble index, timestamp, camera view, and focal-plane setting. This identity is required because the same physical bubble may appear in several rendered observations, while each observation contains a distinct combination of view and defocus information. The resulting supervised sample consists of the crop-level image representation, the known focal-plane depth, the physical regression target, and the associated sample identity.

The DNS depth and diameter values are used only as supervised targets during training and as hidden references during validation and testing. During inference, the model receives the image-derived representation and the known focal-plane information, but not the true bubble depth or diameter.

For the reference synthetic reconstruction benchmark, partitioning is
performed at the DNS-timestamp level. All camera views, focal-plane
settings, and bubble observations generated from a given DNS state
remain within the same training, validation, or test partition. This
prevents alternate optical views of the same underlying flow state
from entering both training and held-out evaluation. The resulting
test therefore measures generalization to DNS states excluded from
the corresponding training partition.

The broad-depth controlled-density stage described in
Sec.~\ref{subsec:broad_depth_learning} serves a complementary purpose.
It expands the supervised distribution in signed focal-plane-relative
depth while controlling projected bubble density. Throughout both
stages, the physical targets, focal-plane convention, optical forward
model, bubble identity, and reconstruction mapping remain unchanged. The stages therefore address different parts of the same inverse problem.

Because the network is trained entirely on ray-traced observations and subsequently evaluated on experimentally acquired shadowgraphs (Sec.~\ref{subsec:experimental_transfer}), the underlying assumption is that the DNS-to-image simulation pipeline is a sufficiently faithful stand-in for the physical imaging process. This is analogous to simulation-to-reality transfer in learning-based robotics and computer vision, where a network trained purely on synthetic renderings is deployed on real sensor data~\citep{tobin2017domainrand}. In that literature, transfer is supported either by explicitly randomizing scene and rendering parameters so that the real domain falls within the training distribution, or, as in the present case, by validating the forward renderer against real measurements so that no such randomization is required. This alternative route, validating a physically based ray-tracing renderer directly against experimental images of optically complex objects, has recently been demonstrated for transparent refractive objects of complex shape~\citep{kalita2026_ray_tracing_transparent}, which supports a validated forward optical model as an established basis for reliable simulation-to-experiment transfer. The forward-model validation reported in Sec.~\ref{sec:forward_framework} therefore serves the same purpose as domain randomization: it makes simulation-trained supervision applicable to experimental images.

The resulting dataset preserves a direct correspondence between the three-dimensional DNS state and the two-dimensional optical evidence. The ray-traced image provides the appearance to be inverted, while the DNS field provides the physical depth and diameter targets. The following subsection describes how bubble-containing image regions are identified and separated into individual instances before crop-level reconstruction.

\subsection{Bubble-evidence localization and instance extraction}
\label{subsec:instance_extraction}

The first stage of the inverse framework converts each full shadowgraphic frame into a set of spatially localized bubble candidates. This operation is nontrivial because the optical support associated with a bubble is not identical to its projected physical support. Depending on focal-plane displacement, local illumination, neighboring bubbles, and optical overlap, a single physical bubble may appear as a sharply bounded dark core, a diffuse halo-dominated structure, an incomplete boundary, or a weak intensity perturbation embedded within a larger connected foreground region. Conversely, several physically distinct bubbles may contribute to one connected image structure. Direct interpretation of connected foreground regions therefore does not provide a reliable bubble count or preserve bubble identity.

The image-analysis stage is consequently divided into two complementary operations. First, semantic segmentation identifies pixels associated with bubble-related optical evidence and separates them from the slowly varying background. Second, instance segmentation resolves the retained foreground into individual bubble bodies. This separation follows the distinction between semantic segmentation, which determines class membership at the pixel level, and instance segmentation, which additionally preserves object identity~\citep{he2017maskrcnn}. Related detection and segmentation strategies have been used in bubble imaging for multiphase-flow applications, including Faster-R-CNN-based localization and shape reconstruction~\citep{haas2020_bubcnn}, bubble detection and tracking during gas evolution~\citep{toyama2025_fasterrcnn}, learned pixel-level segmentation in two-phase flows~\citep{kim2021_universal}, computer-vision-assisted bubble characterization~\citep{nizovtseva2024_bubble_detection_math}, and deep-learning-based segmentation in gas--solid systems~\citep{yang2024_gassolid_bubble_dl}. More broadly, encoder--decoder architectures are well suited to recovering spatially diffuse foreground support, whereas region-based instance segmentation is advantageous when adjacent or overlapping objects must remain distinguishable~\citep{azad2025_losses_segmentation,gfg_unet_2025,sinha2025_maskrcnn_unet}.

For each grayscale image \(I_{t,v,w}(u,v)\), slowly varying illumination is represented by a background estimate \(\mathcal{H}(I_{t,v,w})\). The residual intensity is evaluated together with local contrast and gradient magnitude so that both sharply focused interfaces and weak defocused responses contribute to the semantic evidence field. The bubble-evidence probability is written as

\begin{equation}
P^{\mathrm{bub}}_{t,v,w}
=
\mathcal{S}_{\psi}
\left(
I_{t,v,w},
I_{t,v,w}-\mathcal{H}(I_{t,v,w}),
\mathcal{C}(I_{t,v,w}),
\left\|\nabla I_{t,v,w}\right\|
\right),
\label{eq:bubble_evidence_probability}
\end{equation}

\noindent where \(\mathcal{S}_{\psi}\) denotes the semantic localization operator and \(\mathcal{C}\) denotes a local-contrast operator. The objective of this stage is deliberately broader than final bubble segmentation. The field \(P^{\mathrm{bub}}_{t,v,w}\) identifies regions containing bubble-related optical evidence without yet assigning a unique bubble identity.

Candidate support is obtained by thresholding the semantic probability field, suppressing isolated responses, and grouping the retained foreground into connected regions. These regions identify where bubble-related structure is present but are not interpreted directly as individual bubbles. Instance identity is subsequently recovered through

\begin{equation}
\left\{
M^{\mathrm{inst}}_{i,t,v,w},
q_{i,t,v,w}
\right\}_{i=1}^{\widehat{N}_{t,v,w}}
=
\mathcal{I}_{\omega}
\left(
I_{t,v,w},
P^{\mathrm{bub}}_{t,v,w}
\right),
\label{eq:instance_extraction}
\end{equation}

\noindent where \(\mathcal{I}_{\omega}\) denotes the instance-segmentation operator,
\(M^{\mathrm{inst}}_{i,t,v,w}\) is the mask assigned to bubble instance \(i\),
\(q_{i,t,v,w}\) is the associated detection score, and
\(\widehat{N}_{t,v,w}\) is the number of retained instances. The predicted masks are mapped back to parent-image coordinates, and redundant detections are removed through overlap-based suppression.

\begin{figure}[!htbp]
    \centering
    \includegraphics[
        width=0.9\textwidth,
        keepaspectratio
    ]{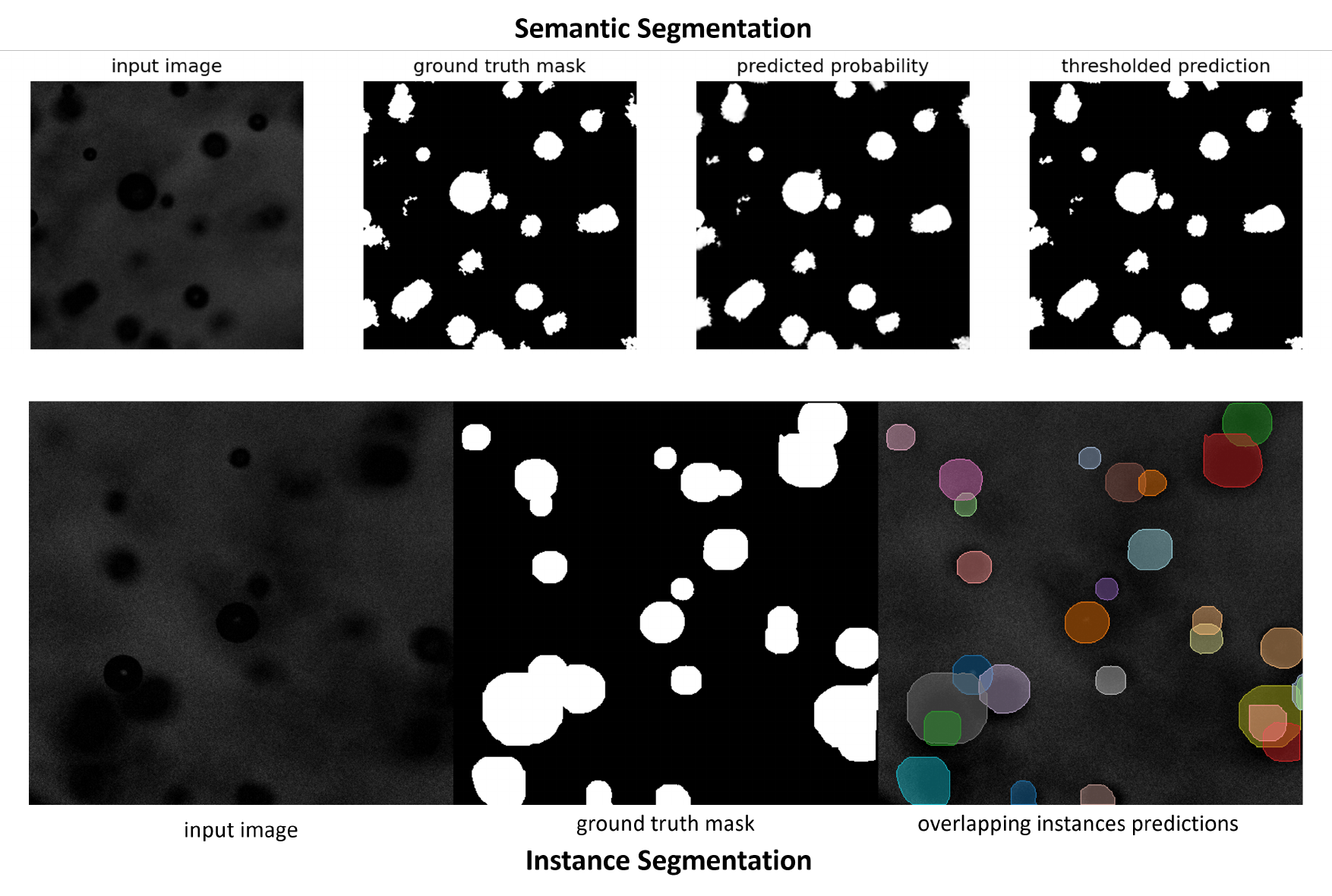}
    \caption{
    Semantic and instance-level decomposition of representative shadowgraphic images.
    The upper row shows semantic segmentation of bubble-related foreground, including the input image, ground-truth foreground support, predicted foreground probability, and thresholded prediction.
    The lower row shows instance segmentation of the retained foreground, where overlapping and touching bubble bodies are assigned separate identities.
    The semantic stage determines where bubble-related optical evidence is present, whereas the instance stage determines how many individual bubbles contribute to that evidence and where their projected supports are located.
    This hierarchical decomposition forms the basis of the bubble-centered inverse reconstruction described in the subsequent sections.
    }
    \label{fig:semantic_instance_segmentation}
\end{figure}

Figure~\ref{fig:semantic_instance_segmentation} illustrates the complementary roles of the two stages. The semantic network first distinguishes bubble-related foreground from background and recovers connected support even when the optical response is weak or diffuse. The subsequent instance stage separates that support into individual projected bubble bodies, including cases in which neighboring bubbles remain connected or partially overlap in the foreground mask. The comparison therefore demonstrates why foreground localization and bubble identity are treated as separate inference tasks in the final pipeline.

For each retained bubble instance, the binary mask provides the image-plane centroid, projected support area, and spatial extent required for crop formation and later mapping to physical lateral coordinates. These quantities are computed geometrically from the predicted mask and are not introduced as separate regression targets.

The complete localization procedure is summarized in Algorithm~\ref{alg:instance_localization}.

\begin{algorithm}[!htbp]
\caption{Semantic localization and bubble-instance extraction}
\label{alg:instance_localization}
\small
\begin{algorithmic}[1]

\Require Grayscale shadowgraph \(I_{t,v,w}\)

\Ensure Bubble instance masks
\(\{M^{\mathrm{inst}}_{i,t,v,w}\}_{i=1}^{\widehat{N}_{t,v,w}}\)

\State Estimate slowly varying background:
\[
B_{t,v,w}
\gets
\mathcal{H}(I_{t,v,w})
\]

\State Form background-corrected residual:
\[
R_{t,v,w}
\gets
I_{t,v,w}-B_{t,v,w}
\]

\State Compute local contrast
\(\mathcal{C}(I_{t,v,w})\)
and gradient magnitude
\(\|\nabla I_{t,v,w}\|\)

\State Evaluate semantic bubble-evidence probability:
\[
P^{\mathrm{bub}}_{t,v,w}
\gets
\mathcal{S}_{\psi}
\left(
I_{t,v,w},
R_{t,v,w},
\mathcal{C}(I_{t,v,w}),
\|\nabla I_{t,v,w}\|
\right)
\]

\State Threshold \(P^{\mathrm{bub}}_{t,v,w}\)

\State Remove isolated or spatially insignificant responses

\State Form connected foreground regions

\For{each retained foreground region}
    \State Extract corresponding grayscale region
    \State Apply instance segmentation
    \State Recover individual bubble masks
\EndFor

\State Map masks to parent-image coordinates

\State Suppress duplicate detections using mask overlap

\State Compute centroid and spatial extent of each retained instance

\State \Return
\(\{M^{\mathrm{inst}}_{i,t,v,w}\}_{i=1}^{\widehat{N}_{t,v,w}}\)

\end{algorithmic}
\end{algorithm}

The resulting instance set provides the spatial basis for the bubble-centered representation used for quantitative depth and diameter reconstruction.

\subsection{Focus-conditioned bubble representation}
\label{subsec:focus_conditioned_representation}

Following instance extraction, the inverse problem is reformulated from full-frame image analysis to bubble-centered reconstruction. This transition is important because the geometric quantities of interest are defined at the level of individual bubbles, while the depth information available to the camera is encoded predominantly through local optical structure. The crop-level representation is therefore constructed to preserve the photometric and geometric cues associated with focal-plane displacement while explicitly retaining the imaging configuration under which each observation was acquired.

Defocus modifies apparent extent, edge sharpness, halo width, grayscale attenuation, and boundary completeness, while overlap determines how much of this information remains directly observable. Classical depth-from-defocus methods infer depth from related variations in blur or point-spread response~\citep{pentland1987newsense}. In the present framework, these signatures are retained directly in the local image representation so that the inverse model can learn their joint dependence on bubble depth and physical size without reducing the observation to a single prescribed blur descriptor.

For each detected bubble instance \(i\), a square grayscale crop is
extracted from the parent shadowgraph and centered on the recovered
instance location,

\begin{equation}
C_{i,t,v,w}
=
\mathcal{P}
\left(
I_{t,v,w},
M^{\mathrm{inst}}_{i,t,v,w}
\right),
\label{eq:instance_grayscale_crop}
\end{equation}

\noindent
where \(\mathcal{P}\) denotes crop extraction and spatial
normalization. The crop is represented on a fixed
\(128\times128\) pixel grid throughout regression training and
inference. Its spatial extent is larger than the nominal instance
support so that the local observation retains halo structure,
attenuated boundaries, grayscale redistribution, and other
defocus-dependent information surrounding the detected bubble.

The segmentation stage provides two useful descriptions of spatial
support. The visible support
\(M^{\mathrm{vis}}_{i,t,v,w}\) denotes the portion of the instance
supported directly by the observed intensity field, while the
estimated physical support
\(M^{\mathrm{phys}}_{i,t,v,w}\) describes the projected extent assigned
to the complete bubble instance. These supports can differ when the
bubble is defocused, partially occluded, or connected optically to a
neighboring bubble.

For clarity, the complete bubble-centered observation may be written
as

\begin{equation}
\mathcal{X}_{i,t,v,w}
=
\left\{
C_{i,t,v,w},
M^{\mathrm{vis}}_{i,t,v,w},
M^{\mathrm{phys}}_{i,t,v,w}
\right\},
\label{eq:crop_support_representation}
\end{equation}

\noindent
where the braces denote the image observation together with its
associated spatial-support information. The support masks define
instance identity, crop placement, spatial extent, and quality of the
localized observation. They are not physical depth or diameter
measurements. In the deployed quantitative regressor, the
convolutional appearance encoder operates on the grayscale component
\(C_{i,t,v,w}\), while the support information remains associated with
the instance throughout localization, crop construction, and
full-frame reconstruction.

This distinction is physically useful. Depth information is carried
by the measured optical field through blur, edge attenuation, halo
structure, apparent extent, and grayscale redistribution. The
segmentation stage determines which local optical structure belongs
to a given bubble. The regression stage then infers the physical
variables from that optical structure without introducing
DNS-derived geometry into the input.

The physical-support estimate is obtained entirely from the
image-analysis sequence. It is not constructed from the
DNS-resolved bubble depth, diameter, or projected geometry.
DNS-resolved quantities remain restricted to supervised targets and
quantitative evaluation. This separation prevents target information
from entering the inference pathway.

The grayscale crop is normalized identically during training,
validation, testing, and experimental inference. The known focal-plane
coordinate is retained as an explicit physical conditioning variable.
This coordinate specifies the optical reference from which the
signed depth displacement is defined. Similar grayscale signatures
observed under different focal settings therefore need not correspond
to the same absolute physical depth.

The focus-conditioned regression representation is

\begin{equation}
\mathbf{r}_{i,t,v,w}
=
\left[
\mathcal{N}_{C}
\left(
C_{i,t,v,w}
\right),
\widetilde{z}_{f}^{(w)}
\right],
\label{eq:focus_conditioned_representation}
\end{equation}

\noindent
where \(\mathcal{N}_{C}\) denotes the fixed grayscale-input
normalization and \(\widetilde{z}_{f}^{(w)}\) is the standardized
coordinate of focal plane \(w\).

Equation~\eqref{eq:focus_conditioned_representation} separates two
sources of information. The grayscale crop contains the local optical
signature associated with bubble size and defocus. The focal-plane
coordinate supplies the known physical reference needed to interpret
that signature as a signed displacement. The segmentation-derived
supports remain linked to the same instance and define where that
optical evidence is localized in the parent image.

\subsection{Multitask depth and diameter reconstruction}
\label{subsec:multitask_inverse_model}

\begin{figure}[!htbp]
    \centering
    \includegraphics[
        width=0.9\textwidth,
        keepaspectratio
    ]{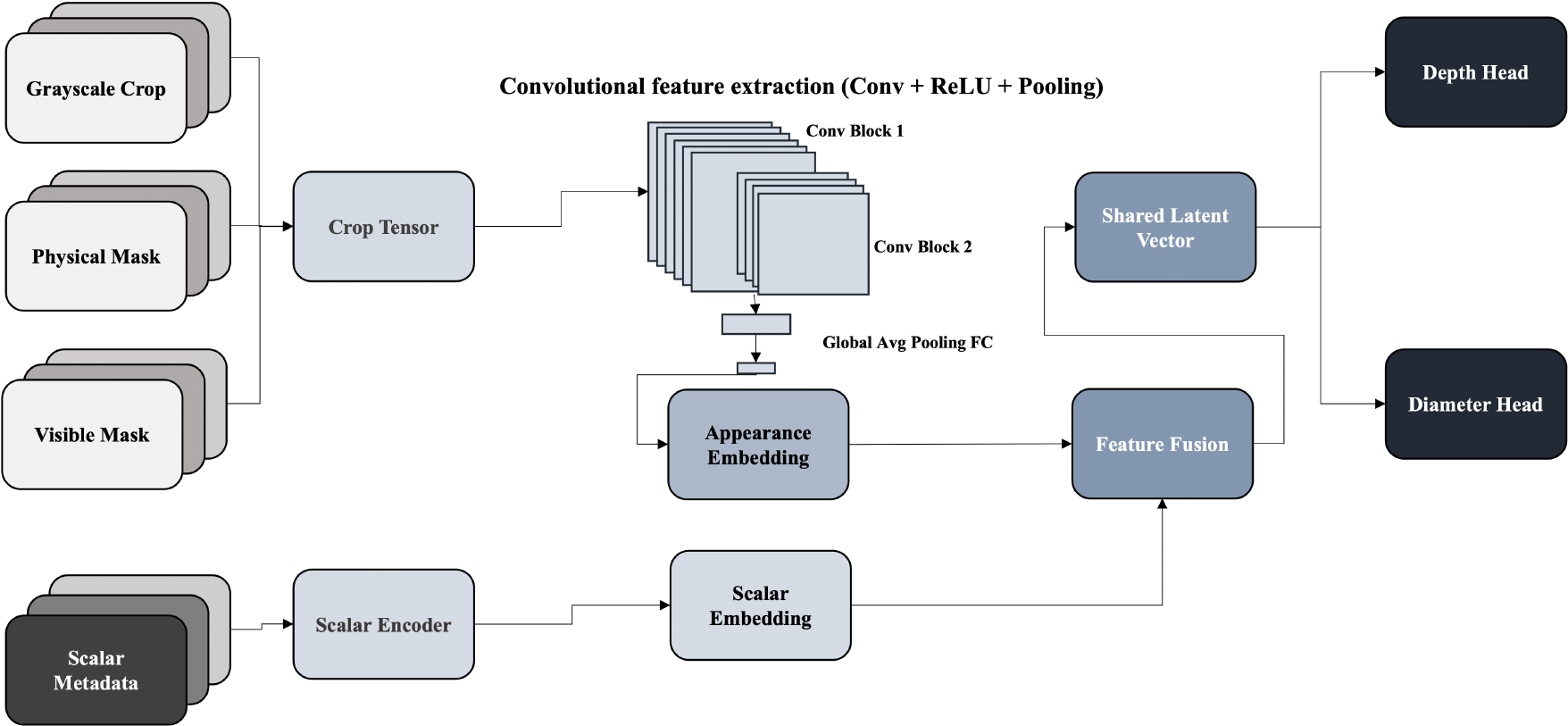}
    \caption{
    Focus-conditioned multitask architecture for bubble-level depth
    and diameter reconstruction. The grayscale crop is associated
    with visible and estimated physical support obtained during
    instance extraction. These supports define instance localization,
    crop geometry, and spatial correspondence. The deployed
    convolutional appearance encoder operates on the normalized
    grayscale crop. Its learned representation is combined with the
    known focal-plane coordinate before joint prediction of signed
    focal-plane-relative depth and physical-equivalent diameter.
    Absolute depth is reconstructed using the known focal-plane
    position.
    }
    \label{fig:inverse_model_architecture}
\end{figure}

The focus-conditioned representation is supplied to a multitask
regression network that jointly estimates signed
focal-plane-relative depth and physical-equivalent diameter
(Fig.~\ref{fig:inverse_model_architecture}). Joint prediction is
appropriate because the dominant optical cues are not independent
functions of the two target variables. Apparent extent, edge
attenuation, halo development, and grayscale contrast depend on both
physical bubble size and displacement from focus. A common appearance
representation therefore allows the two outputs to be inferred from
the same optical evidence.

The deployed image encoder receives a single
\(128\times128\) grayscale crop. It consists of four strided
convolutional blocks with 24, 48, 96, and 160 output channels. The
first convolution uses a \(5\times5\) kernel, while the remaining
three use \(3\times3\) kernels. All four convolutions use stride 2.
Each convolution is followed by batch normalization and a ReLU
activation. Adaptive global-average pooling reduces the final feature
map to a 160-component appearance vector.

Let
\(\mathbf{h}^{\mathrm{img}}_{i,t,v,w}\)
denote this learned representation, $\mathbf{h}^{\mathrm{img}}_{i,t,v,w}=\mathcal{E}_{\theta_e}[\mathcal{N}_{C}(C_{i,t,v,w})]$.

The standardized focal-plane coordinate is concatenated with the
160-component image representation, producing the
focus-conditioned vector $\mathbf{h}^{\mathrm{f}}_{i,t,v,w}=[\mathbf{h}^{\mathrm{img}}_{i,t,v,w},\widetilde{z}_{f}^{(w)}]\in\mathbb{R}^{161}$.

The regression head maps this vector through fully connected layers
of width 128 and 64. A dropout probability of 0.10 is applied after
the first fully connected layer. The final two outputs are the
normalized signed relative depth and physical-equivalent diameter,

\begin{equation}
\left[
\widehat{\Delta z}_{i,n},
\widehat{d}_{i,n}
\right]
=
\mathcal{G}_{\theta}
\left(
\mathbf{h}^{\mathrm{f}}_{i,t,v,w}
\right),
\label{eq:multitask_inverse_mapping}
\end{equation}

\noindent
where the subscript \(n\) denotes target normalization using
statistics computed from the training partition. The normalized
outputs are transformed back to physical units before reconstruction,
and absolute depth is obtained from
Eq.~\eqref{eq:relative_absolute_depth}.

The shared formulation is physically appropriate because an increase
in apparent bubble extent may be produced by a larger physical
diameter, a change in focal-plane displacement, or both. Likewise,
halo development and edge attenuation carry depth information while
remaining dependent on bubble size and local optical overlap. Joint
regression allows these coupled signatures to be interpreted together
with each image feature informing both physical variables.

For a mini-batch of \(B\) bubble instances, the two normalized
regression targets are optimized using a weighted Smooth-\(L_1\)
objective,

\begin{equation}
\mathcal{L}_{\mathrm{multi}}
=
\frac{1}{B}
\sum_{i=1}^{B}
w_i
\frac{
\ell_{\mathrm{SL1}}
\left(
\widehat{\Delta z}_{i,n},
\Delta z_{i,n}
\right)
+
\ell_{\mathrm{SL1}}
\left(
\widehat{d}_{i,n},
d_{i,n}
\right)
}{2},
\label{eq:multitask_loss}
\end{equation}

\noindent
where \(\ell_{\mathrm{SL1}}\) denotes the Smooth-\(L_1\) loss.
The depth-dependent sample weight is

\begin{equation}
w_i
=
1+
\min
\left[
3,\,
\frac{
\left|\Delta z_i\right|
}{
Q_{75}\left(\left|\Delta z\right|\right)
}
\right],
\label{eq:depth_weight}
\end{equation}

\noindent
where
\(Q_{75}(|\Delta z|)\)
is the 75th percentile of the absolute relative-depth distribution
evaluated from the training partition. Thus,
\(1\leq w_i\leq4\).
The weighting increases the contribution of strongly defocused
examples without changing their target values. This is useful because
their optical signatures are weaker and less numerous than those
close to focus.

The broad-depth crop database contains 138,240 labeled bubble
instances. For optimization and checkpoint selection, 110,592 crops
are used for training and 27,648 for validation, corresponding to an
80/20 randomized partition with random seed 42. Training is performed
for 40 epochs with a batch size of 64 using AdamW with learning rate
\(2\times10^{-4}\) and weight decay \(10^{-4}\). No learning-rate
scheduler is applied. The checkpoint with minimum validation loss is
retained. The selected checkpoint occurs at epoch 17.

\begin{table}[!htbp]
\caption{
Architecture and optimization configuration of the
focus-conditioned broad-depth crop regressor.
}
\label{tab:inverse_model_configuration}
\centering
\small
\begin{tabular}{ll}
Quantity & Configuration \\
\hline
Crop size & \(128\times128\) pixels \\
Regression image channels & 1 grayscale channel \\
Conv. block 1 & \(1\rightarrow24,\;5\times5,\;s=2\) \\
Conv. block 2 & \(24\rightarrow48,\;3\times3,\;s=2\) \\
Conv. block 3 & \(48\rightarrow96,\;3\times3,\;s=2\) \\
Conv. block 4 & \(96\rightarrow160,\;3\times3,\;s=2\) \\
Conv. normalization & Batch normalization \\
Conv. activation & ReLU \\
Pooling & adaptive global average \\
Image embedding & 160 components \\
Focus conditioning & standardized \(z_f\) \\
Fused representation & 161 components \\
Regression head & \(161\rightarrow128\rightarrow64\rightarrow2\) \\
Dropout & 0.10 \\
Optimizer & AdamW \\
Learning rate & \(2\times10^{-4}\) \\
Weight decay & \(10^{-4}\) \\
Batch size & 64 \\
Training epochs & 40 \\
Checkpoint selection & minimum validation loss \\
Selected epoch & 17 \\
\end{tabular}
\end{table}

The architecture and training configuration in
Table~\ref{tab:inverse_model_configuration} define the broad-depth
focus-conditioned regressor used for depth-resolved reconstruction.
The timestamp-disjoint synthetic evaluation described in
Sec.~\ref{subsec:closed_loop_validation} provides the complementary
generalization test under independently held-out DNS states.
Both evaluations use the same physical depth convention, diameter
definition, forward optical model, bubble-centered inverse
formulation, and three-dimensional reconstruction mapping.

The resulting model provides bubble-level depth and diameter estimates
over the optical conditions represented in the supervised data.
Extending the reconstruction across the measurement depth requires
training examples that retain distant defocus signatures while
preserving unambiguous correspondence between rendered bubbles and
their physical labels. The following subsection describes the
controlled-density strategy used to construct this broad-depth
supervision.

The resulting model provides bubble-level depth and diameter estimates over the optical conditions represented in the supervised data. Extending this reconstruction across a broader depth interval requires training examples that preserve distant defocus signatures while maintaining unambiguous correspondence between rendered bubbles and physical labels. The following subsection describes the controlled-density strategy used to construct this broad-depth supervision.

\subsection{Broad-depth controlled-density learning}
\label{subsec:broad_depth_learning}

The initial inverse model was trained using bubbles located within relatively narrow depth intervals around each focal plane. This provided a well-constrained setting for learning the local relationship between optical appearance, signed displacement from focus, and physical diameter. However, reconstruction over the complete measurement depth requires supervised examples that include both near-focus and strongly defocused bubbles.

Rendering the entire DNS population across a broad depth interval creates a second difficulty. As the projected number density increases, focused cores, halos, and diffuse shadow structures overlap in the image plane. Weak evidence from distant bubbles may then be obscured by nearby bubbles, reducing the correspondence between an individual image signature and its DNS label. Broad depth coverage must therefore be introduced without allowing excessive projected crowding to dominate the supervised signal.

To satisfy both requirements, the training data are constructed using a broad-depth controlled-density strategy, which combines stratified sampling over signed depth with explicit control of the rendered instance density. Depth-stratified sampling is analogous in spirit to curriculum and balanced-sampling strategies used elsewhere in supervised learning, in which the composition of training batches is deliberately controlled, so that the learner is exposed to a prescribed range of task difficulty and class or condition frequency~\citep{bengio2009curriculum}. Here, the controlled dimension is signed focal-plane-relative depth, and the explicit density control additionally regulates the amount of projected bubble-to-bubble overlap admitted into each rendered frame. For focal-plane setting \(w\), the admissible signed-depth interval is divided into \(K\) strata. Bubbles are grouped according to their focal-plane-relative depth and a prescribed number is selected from each stratum,

\begin{equation}
\begin{aligned}
\mathcal{D}_{k} &= \left[\delta_{k-1}, \delta_k\right), \qquad \mathcal{B}_{t,w}^{(k)} = \left\{\mathbf{g}_{i,t}\in\mathcal{B}_{t} : z_{i,t}-z_f^{(w)} \in \mathcal{D}_{k}\right\}, \\
\widetilde{\mathcal{B}}_{t,v,w} &= \bigcup_{k=1}^{K}\widetilde{\mathcal{B}}_{t,v,w}^{(k)}, \qquad \widetilde{\mathcal{B}}_{t,v,w}^{(k)} \subseteq \mathcal{B}_{t,w}^{(k)}, \qquad \left|\widetilde{\mathcal{B}}_{t,v,w}^{(k)}\right| = n_k .
\end{aligned}
\label{eq:bdcd_sampling}
\end{equation}

Here, \(n_k\) controls the contribution of depth stratum \(k\) to the rendered frame. The resulting sampled population includes focused, near-defocused, and far-defocused bubbles on both sides of the focal plane while preventing naturally dense depth regions from dominating the training distribution.

The total number of selected bubbles determines the controlled-density condition. Different density conditions expose the model to varying degrees of isolation and optical overlap while preserving the same physical depth domain and focal-plane geometry. This allows the model to learn from moderate interactions between bubble signatures without eliminating the defocus information required for depth reconstruction.

Only the selected bubbles are supplied to the forward optical model,

\begin{equation}
\begin{aligned}
I^{\mathrm{BDCD}}_{t,v,w} &= \mathcal{R}\left(\widetilde{\mathcal{B}}_{t,v,w}, v, z_f^{(w)}\right), \qquad \mathcal{B}^{\mathrm{render}}_{t,v,w} = \mathcal{B}^{\mathrm{label}}_{t,v,w} = \widetilde{\mathcal{B}}_{t,v,w}.
\end{aligned}
\label{eq:bdcd_render_label_contract}
\end{equation}

This rendered-label identity is essential for preserving bubble-level supervision. No unsampled bubble contributes optical structure to the image, and no retained label lacks corresponding rendered evidence. The per-frame sampling procedure of Eq.~\eqref{eq:bdcd_sampling} is used to construct \(\widetilde{\mathcal{B}}_{t,v,w}\) from the full DNS population before rendering.

\begin{algorithm}[!htbp]
\caption{Construction of the focus-conditioned bubble representation}
\label{alg:focus_conditioned_representation}
\small
\begin{algorithmic}[1]

\Require
grayscale image \(I_{t,v,w}\),
instance mask \(M^{\mathrm{inst}}_{i,t,v,w}\),
and focal-plane coordinate \(z_f^{(w)}\)

\Ensure
focus-conditioned regression representation
\(\mathbf{r}_{i,t,v,w}\)

\State compute the image-plane centroid and spatial extent of
\(M^{\mathrm{inst}}_{i,t,v,w}\)

\State extract a fixed \(128\times128\) grayscale crop:
\[
C_{i,t,v,w}
\gets
\mathcal{P}
\left(
I_{t,v,w},
M^{\mathrm{inst}}_{i,t,v,w}
\right)
\]

\State determine visible support
\(M^{\mathrm{vis}}_{i,t,v,w}\)

\State estimate physical support
\(M^{\mathrm{phys}}_{i,t,v,w}\)

\State associate
\(
\{C_{i,t,v,w},
M^{\mathrm{vis}}_{i,t,v,w},
M^{\mathrm{phys}}_{i,t,v,w}\}
\)
with the retained bubble instance

\State normalize grayscale crop:
\[
\widetilde{C}_{i,t,v,w}
\gets
\mathcal{N}_{C}
\left(
C_{i,t,v,w}
\right)
\]

\State normalize focal-plane coordinate:
\[
\widetilde{z}_{f}^{(w)}
\gets
\mathcal{N}_{z}
\left(
z_f^{(w)}
\right)
\]

\State form regression representation:
\[
\mathbf{r}_{i,t,v,w}
\gets
\left[
\widetilde{C}_{i,t,v,w},
\widetilde{z}_{f}^{(w)}
\right]
\]

\State
\Return
\(\mathbf{r}_{i,t,v,w}\)

\end{algorithmic}
\end{algorithm}

The controlled-density procedure changes only which DNS bubbles coexist within a training frame. It does not alter the position, diameter, or optical response of any selected bubble. Each bubble retains its original DNS-resolved state and is rendered using the same forward optical model described previously.

Consequently, controlled-density sampling changes the statistical
composition of the rendered observation but not the physical inverse
mapping assigned to any selected bubble. The target pair
\((\Delta z_i,d_i)\), the focal-plane reference, and the ray-traced
optical response remain those of the original DNS-resolved bubble.
The procedure therefore broadens the range of observable defocus
conditions without introducing synthetic changes to the supervised
physical labels.

The resulting training distribution spans multiple DNS timestamps, camera views, focal-plane settings, signed-depth strata, and density conditions. It therefore extends the supervised depth range while retaining sufficient optical separability for bubble-level learning. The trained model can then be applied to all retained instances in a full image and assembled into a three-dimensional bubble population.

\subsection{Full-frame three-dimensional reconstruction}
\label{subsec:full_frame_reconstruction}

Following bubble-level inference, each prediction is mapped from the crop coordinate system to the parent image and then expressed in physical coordinates. This step converts the local depth and diameter estimates into a frame-resolved three-dimensional bubble population.

The image-plane centroid of each retained instance is obtained from its segmentation mask. The corresponding lateral coordinates are recovered using the calibrated image-to-object transformation associated with camera view \(v\),

\begin{equation}
\begin{aligned}
\widehat{x}_{i,t,v,w}
&=
\mathcal{T}_{x}
\left(
\widehat{u}_{i,t,v,w},
\widehat{v}_{i,t,v,w},
v
\right),
\\
\widehat{y}_{i,t,v,w}
&=
\mathcal{T}_{y}
\left(
\widehat{u}_{i,t,v,w},
\widehat{v}_{i,t,v,w},
v
\right).
\end{aligned}
\label{eq:lateral_coordinate_mapping}
\end{equation}

For a spatially uniform calibration aligned with the physical coordinate system, the transformations reduce to multiplication by the calibrated physical pixel scale after reference-origin correction. The absolute depth is obtained from the known focal-plane location and the predicted relative depth using Eq.~\eqref{eq:relative_absolute_depth}. Together with the predicted diameter and instance confidence, these quantities define the reconstructed bubble state introduced in Eq.~\eqref{eq:inverse_population}.

Each reconstructed bubble remains linked to its source image, segmentation mask, crop location, focal-plane setting, and confidence score. Preserving this identity enables both bubble-level comparison and frame-level population analysis without losing the correspondence between the optical evidence and the reconstructed physical state.

For visualization, the reconstructed bubble may be represented as a sphere centered at
\(\left(\widehat{x}_{i,t,v,w},\widehat{y}_{i,t,v,w},\widehat{z}_{i,t,v,w}\right)\)
with radius \(\widehat{d}_{i,t,v,w}/2\). This representation is used only to visualize the predicted position and physical-equivalent diameter. Quantitative evaluation is performed using the continuous depth and diameter predictions.

For depth-resolved population analysis, reconstructed bubbles are assigned to physical sampling slabs centered at \(z_k\). The reconstructed population associated with slab \(k\) is

\begin{equation}
\widehat{\mathcal{B}}_{k}
=
\left\{
\widehat{\mathbf{b}}_{i,t,v,w}
:
\left|
\widehat{z}_{i,t,v,w}
-
z_k
\right|
\leq
\frac{W_k}{2}
\right\},
\label{eq:plane_reconstructed_population}
\end{equation}

\noindent where \(W_k\) is the physical slab width. The corresponding diameter population is obtained directly from the predicted diameters of the bubbles contained in \(\widehat{\mathcal{B}}_{k}\).

This slab-based representation connects bubble-level predictions to depth-dependent flow statistics, including bubble-size probability distributions and characteristic diameters. Predictions outside the physically admissible depth or diameter range are excluded from population-level analysis but are not otherwise modified.

The reconstructed three-dimensional population provides the common output for both validation settings. Under DNS-controlled conditions, predicted bubble positions and diameters can be compared directly with the known physical field. Under experimental conditions, the reconstructed population is evaluated using depth-resolved size distributions and aggregate bubble statistics. This common representation is what makes the two validation settings directly comparable: both are evaluated as instances of the same reconstructed population defined in Eq.~\eqref{eq:inverse_population}, differing only in whether an independent three-dimensional reference is available for direct bubble-level comparison. The following section describes the closed-loop DNS validation of the complete reconstruction sequence and its extension to experimental transfer.

\FloatBarrier

\section{Validation methodology}
\label{sec:validation_methodology}

\subsection{Closed-loop DNS-to-image-to-reconstruction validation}
\label{subsec:closed_loop_validation}

The complete inverse pipeline is first evaluated under controlled DNS conditions, where the three-dimensional bubble field used to generate each optical image is known. This closed-loop setting enables direct comparison between the reconstructed population and the physical state supplied to the forward optical model.

For each selected timestamp, camera view, and focal-plane setting, the validation sequence is

\begin{equation}
\mathcal{B}_{t}
\overset{\mathcal{R}}{\longrightarrow}
I_{t,v,w}
\overset{\mathcal{S}}{\longrightarrow}
\mathcal{M}_{t,v,w}
\overset{\mathcal{G}_{\theta}}{\longrightarrow}
\widehat{\mathcal{B}}_{t,v,w},
\label{eq:closed_loop_validation_chain}
\end{equation}

\noindent where \(\mathcal{R}\) denotes the forward ray-tracing operator, \(\mathcal{S}\) represents bubble localization and instance extraction, and \(\mathcal{G}_{\theta}\) denotes the focus-conditioned inverse model. The reconstructed population is assembled using the lateral coordinate mapping and absolute-depth relation defined previously.

This validation evaluates the complete reconstruction sequence. Differences between the DNS and reconstructed populations may arise from missed detections, instance separation, crop formation, depth regression, diameter regression, or mapping of bubble-level predictions back to the full frame.

Reconstructed instances are linked to their corresponding DNS bubbles using the retained sample identities. Bubble-level errors are calculated only after inference has been completed, and the DNS-resolved depth and diameter are not supplied to the model during prediction.

Reconstruction performance is quantified using depth mean absolute error, diameter mean absolute error, diameter mean absolute percentage error, normalized depth accuracy, and diameter accuracy. The absolute-error measures retain direct interpretation in physical units, while the normalized accuracy measures provide compact summaries across the rendered depth and diameter ranges.

\begin{figure}[!htbp]
    \centering
    \includegraphics[
        width=0.95\textwidth,
        keepaspectratio
    ]{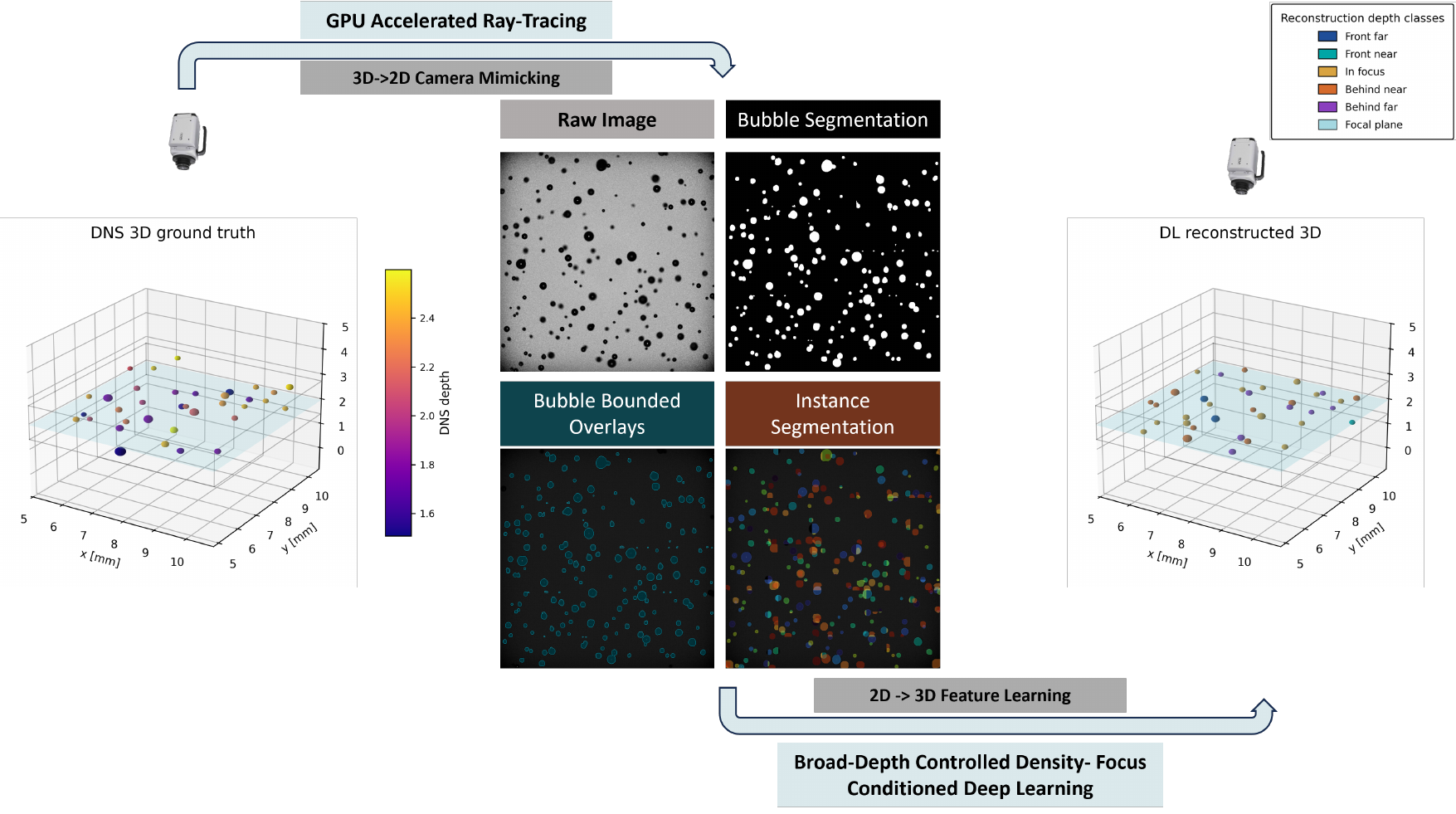}
    \caption{
    Closed-loop DNS-to-image-to-reconstruction validation.
    A DNS-resolved three-dimensional bubble field is converted into a grayscale shadowgraphic image using the forward ray-tracing model.
    Bubble instances are localized and supplied to the focus-conditioned inverse model, which predicts focal-plane-relative depth and physical-equivalent diameter.
    The resulting bubble-level estimates are then assembled into the reconstructed three-dimensional population.
    }
    \label{fig:3d_2d_3d_reconstruction}
\end{figure}

Figure~\ref{fig:3d_2d_3d_reconstruction} compares the DNS-resolved bubble field, its ray-traced image, the extracted bubble instances, and the corresponding reconstruction for the same timestamp and optical configuration. This comparison determines whether the spatial and size information contained in the DNS field is preserved through the complete forward and inverse sequence.

Depth reconstruction performance is quantified using the mean absolute error, $\mathrm{MAE}_{z}=N^{-1}\sum_{i=1}^{N}\left|\widehat{z}_{i}-z_i\right|$.

The normalized depth accuracy is defined relative to the full physical depth range represented by the rendered imaging domain as $\mathrm{Acc}_{z}=100\left[1-\mathrm{MAE}_{z}/\left(z_{\max}^{\mathrm{domain}}-z_{\min}^{\mathrm{domain}}\right)\right]$.

Here, $N$ is the number of evaluated bubble instances, $\widehat{z}_{i}$ is the reconstructed depth, and $z_i$ is the corresponding DNS-resolved depth. The quantity $z_{\max}^{\mathrm{domain}}-z_{\min}^{\mathrm{domain}}$ denotes the complete physical depth interval represented in the rendered dataset.

Quantitatively, the crop-level regressor achieved a normalized depth accuracy of \(\sim97.6\%\) on the held-out synthetic test partition, indicating that the signed focal-plane-relative depth is recovered consistently across DNS timestamps that were not seen during training.
Applying the complete closed-loop sequence of Eq.~\eqref{eq:closed_loop_validation_chain} to a same-frame test containing \(180\) DNS-resolved bubbles yielded a diameter mean absolute percentage error below \(9.71\%\), corresponding to a diameter accuracy of \(\sim90.29\%\). The small reduction from the crop-level depth accuracy to the closed-loop diameter accuracy is consistent with the closed-loop test accumulating error from localization and instance separation in addition to the depth and diameter regression heads, as anticipated in the discussion of Eq.~\eqref{eq:closed_loop_validation_chain} above. Together, these results indicate that bubble-level geometric information is retained through the complete three-dimensional-to-two-dimensional-to-three-dimensional cycle.

The DNS-controlled analysis establishes bubble-level reconstruction performance under known physical ground truth. Experimental images do not provide an independent three-dimensional reference for every detected bubble. The same inference framework is therefore evaluated experimentally through depth-resolved population statistics and bubble-size distributions.

\subsection{Transfer to experimental shadowgraph images}
\label{subsec:experimental_transfer}

Following DNS-controlled validation, the trained inverse framework is applied to experimental shadowgraph images acquired at known focal-plane positions. The experimental images are processed using the same localization, instance-extraction, crop-formation, and focus-conditioned regression sequence established for the rendered data.

Before inference, each experimental frame is subjected to a fixed intensity-normalization procedure to reduce differences associated with illumination nonuniformity, sensor response, background structure, optical interfaces, and image noise. This preprocessing is applied consistently across all frames and preserves the local grayscale structures associated with focused boundaries, defocused halos, contrast transitions, and overlapping bubble signatures.

For an experimental frame \(I^{\mathrm{exp}}_{m,f}\), acquired under focal condition \(f\), the inverse model predicts focal-plane-relative depth and physical-equivalent diameter. The absolute depth is then reconstructed by combining the predicted relative depth with the known focal-plane position, following the formulation introduced previously.

The physical focal-plane locations used in the experimental analysis are
\(7.20\ \mathrm{mm}\) for RADIAL CENTER,
\(5.07\ \mathrm{mm}\) for RADIAL NEAR CENTER,
\(2.93\ \mathrm{mm}\) for RADIAL NEAR WALL, and
\(0.80\ \mathrm{mm}\) for RADIAL WALL.
These locations correspond to the radial measurement geometry of the
experimental facility documented by Kumar et al.~\citep{kumar2026bubble},
where the camera is translated across the duct cross-section to obtain
plane-resolved bubble measurements from the wall toward the channel
centerline. In the present analysis, these coordinates define both the
focal-plane metadata supplied to the inverse model and the physical
depth locations used for plane-resolved reconstruction. Application of
the trained inverse model to the experimental frames acquired at these
four planes yielded \(142\), \(971\), \(859\), and \(204\) reconstructed
bubbles at RADIAL CENTER, RADIAL NEAR CENTER, RADIAL NEAR WALL, and
RADIAL WALL, respectively. These populations are used for the
plane-resolved statistical comparisons presented below.

The continuous reconstructed depths are assigned to physical sampling slabs using the criterion defined in Eq.~\eqref{eq:plane_reconstructed_population}. A slab width of \(1.5\ \mathrm{mm}\) is used for the radial comparison. The diameter population associated with each plane is then formed from the predicted diameters of bubbles whose reconstructed depths fall within the corresponding slab.

Unlike the DNS-controlled images, the experimental data do not provide independently known three-dimensional depth and physical diameter for every detected bubble. Direct bubble-to-bubble error evaluation is therefore unavailable for the complete experimental population.

Experimental validation is performed at the population level. For each physical plane, the reconstructed diameter distribution is compared with the corresponding experimental reference distribution obtained under the in-focus measurement condition for that plane. This comparison assesses whether the reconstruction preserves the dominant size range, distribution width, and large-bubble tail observed experimentally.

The following subsection defines the distribution-based and characteristic-diameter metrics used for this population-level assessment.

\subsection{Distribution-level validation and recoverability analysis}
\label{subsec:distribution_validation}
Because bubble-resolved experimental ground truth is unavailable, the reconstructed populations are evaluated using depth-resolved bubble-size statistics. For each physical plane \(k\), the predicted diameter population is compared with the corresponding in-focus experimental reference using common diameter-bin edges.

Let \(\widehat{n}_{k,j}\) and \(n^{\mathrm{exp}}_{k,j}\) denote the reconstructed and experimental bubble counts in diameter bin \(j\). The normalized bin probabilities and their distribution overlap are defined as

\begin{equation}
\begin{aligned}
\widehat{p}_{k,j} &= \frac{\widehat{n}_{k,j}}{\sum_{\ell=1}^{J}\widehat{n}_{k,\ell}}, \\
p^{\mathrm{exp}}_{k,j} &= \frac{n^{\mathrm{exp}}_{k,j}}{\sum_{\ell=1}^{J}n^{\mathrm{exp}}_{k,\ell}}, \\
O_k &= \sum_{j=1}^{J}\min\left(\widehat{p}_{k,j}, p^{\mathrm{exp}}_{k,j}\right), \qquad E_{\mathrm{PDF},k} = 100\left(1-O_k\right).
\end{aligned}
\label{eq:distribution_overlap}
\end{equation}

Here, \(O_k=1\) indicates identical normalized distributions, while \(O_k=0\) indicates no common probability support. The corresponding error \(E_{\mathrm{PDF},k}\) evaluates agreement across the complete bubble-size distribution, including its dominant range, width, and tails.

\begin{figure}[!htbp]
    \centering
    \includegraphics[
        width=0.85\textwidth,
        keepaspectratio
    ]{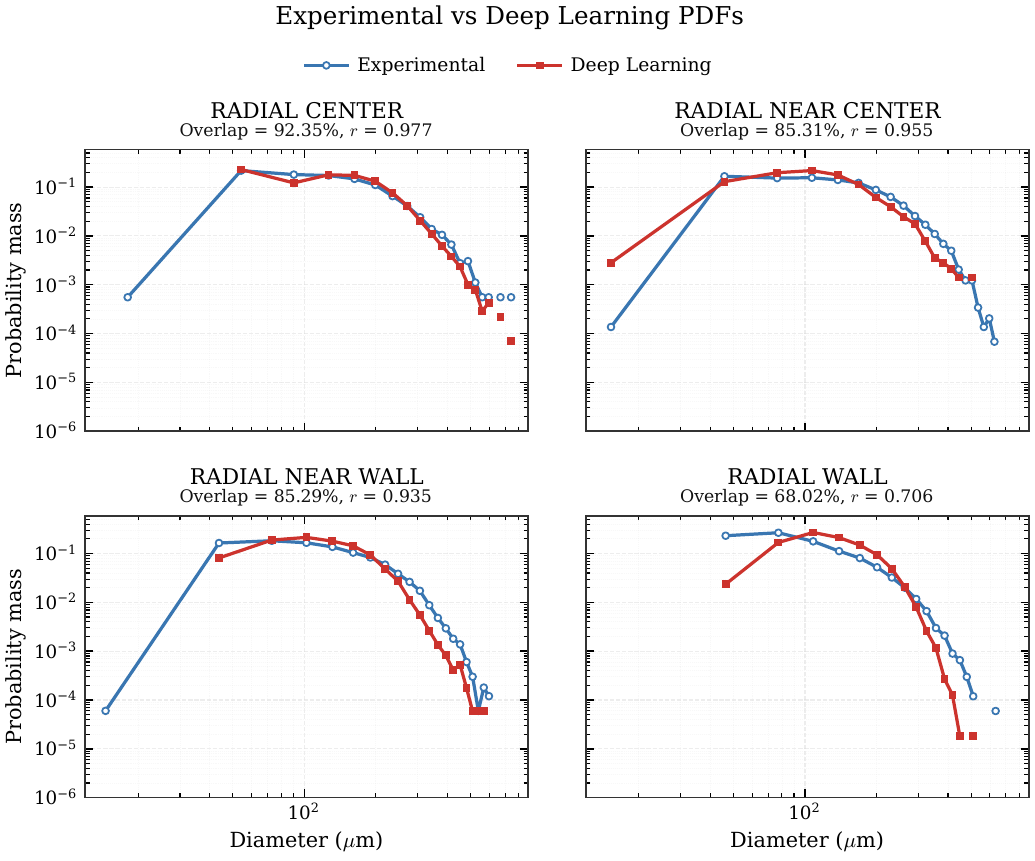}
    \caption{
    Comparison of experimental and reconstructed bubble-size probability distributions across the four radial validation planes.
    Each panel shows the in-focus experimental reference and the corresponding reconstructed distribution using common diameter-bin edges and normalized probability per bin.
    }
    \label{fig:matlab_style_exp_dl_pdf_all_planes}
\end{figure}

Across the four radial planes shown in Fig.~\ref{fig:matlab_style_exp_dl_pdf_all_planes}, the reconstructed and experimental bubble-size distributions exhibited PDF-overlap similarities \(O_k\) of \(\sim92\%\), \(\sim85\%\), \(\sim85\%\), and \(\sim68\%\) at RADIAL CENTER, RADIAL NEAR CENTER, RADIAL NEAR WALL, and RADIAL WALL, respectively, with corresponding log-PDF correlations of \(0.7161\), \(0.8183\), \(0.8718\), and \(0.8518\). For the RADIAL CENTER reference condition, for which \(11{,}679\) experimentally detected bubbles established the reference distribution, the reconstructed and experimental center-plane distributions exhibited a similarity of \(\sim86.37\%\) with a correlation of \(0.9278\). Taken together, these metrics indicate that the inverse model preserves the principal structure and radial variation of the experimental bubble-size distributions across the measurement depth, while the lower overlap at RADIAL WALL already signals the reduced agreement expected far from the focal reference plane, examined quantitatively below.

The full-distribution comparison is complemented by two characteristic diameter measures. The Sauter mean diameter characterizes the size-weighted bulk population, while the \(99.8\)-th percentile diameter characterizes the upper tail,

\begin{equation}
\begin{aligned}
D_{32} &= \frac{\sum_{i=1}^{N}d_i^3}{\sum_{i=1}^{N}d_i^2}, \qquad D_{99.8,k} = F_k^{-1}(0.998), \\
E_{32,k} &= 100\frac{\left|\widehat{D}_{32,k}-D_{32,k}^{\mathrm{exp}}\right|}{D_{32,k}^{\mathrm{exp}}}, \\
E_{99.8,k} &= 100\frac{\left|\widehat{D}_{99.8,k}-D_{99.8,k}^{\mathrm{exp}}\right|}{D_{99.8,k}^{\mathrm{exp}}}.
\end{aligned}
\label{eq:characteristic_diameter_metrics}
\end{equation}

The three diagnostics are interpreted jointly. The PDF error measures agreement across the complete population, \(E_{32,k}\) emphasizes the size-weighted bulk distribution, and \(E_{99.8,k}\) measures preservation of the large-bubble tail. In particular, a small \(D_{32}\) error does not necessarily imply close distributional agreement because compensating changes in the weighted moments may preserve their ratio.

Recoverability is evaluated as a function of the distance between physical plane \(k\) and the focal plane, $r_k=\left|z_k-z_f\right|$.

When a single summary measure is required, the three plane-resolved errors are combined as

\begin{equation}
E_{\mathrm{rec},k}
=
w_{\mathrm{PDF}}E_{\mathrm{PDF},k}
+
w_{32}E_{32,k}
+
w_{99.8}E_{99.8,k},
\qquad
w_{\mathrm{PDF}}+w_{32}+w_{99.8}=1.
\label{eq:recoverability}
\end{equation}

\noindent
where the nonnegative weights are fixed consistently across all planes.
The composite measure is reported together with its individual components
so that agreement in one statistic does not conceal degradation in another.

\begin{figure}[!htbp]
    \centering
    \includegraphics[
        width=0.85\textwidth,
        keepaspectratio
    ]{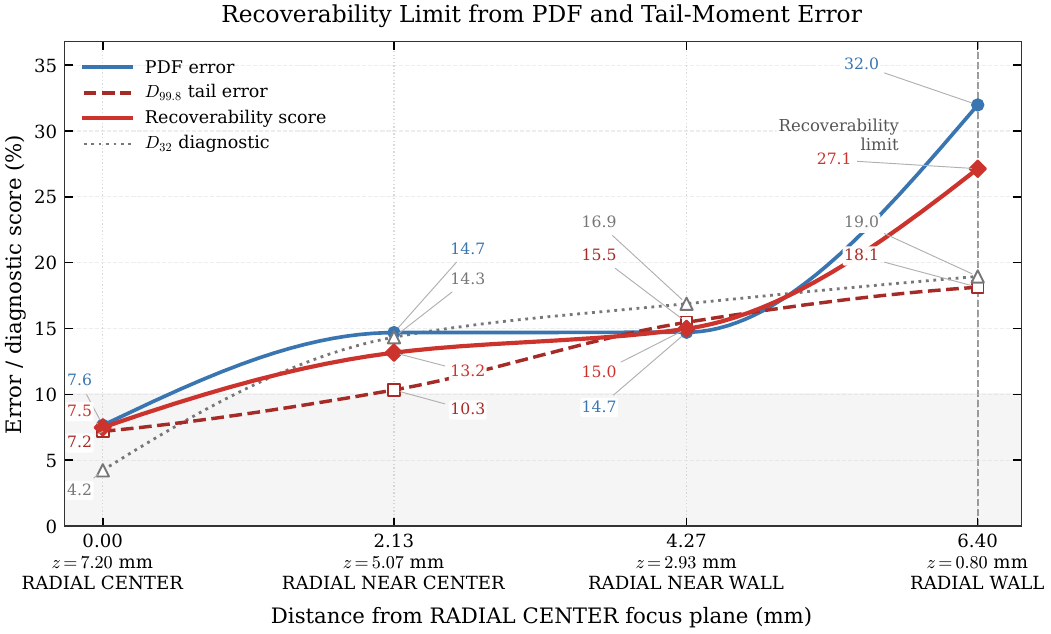}
    \caption{
    Recoverability of depth-resolved bubble-size statistics as a function of distance from the reference focal plane.
    The analysis includes the full-distribution error, \(D_{32}\) error, and \(D_{99.8}\) error at each radial plane.
    Their variation with focal-plane distance identifies the practical depth range over which a single observation retains sufficient information for population reconstruction.
    }
    \label{fig:recoverability_limit_single_plane}
\end{figure}

The variation of these errors with distance from the focal plane is summarized in Fig.~\ref{fig:recoverability_limit_single_plane}.

Near the focal plane, stronger boundary definition, contrast, and apparent-size information support closer agreement with the experimental reference. With increasing focal-plane distance, diffuse optical support, overlap, and reduced uniqueness of the depth-diameter relation increase reconstruction uncertainty. At RADIAL CENTER, the plane coincident with its own focal setting (\(r_k=0\)), the full-distribution error remained below \(7.65\%\), the \(D_{99.8}\) error remained below \(7.18\%\), and the composite recoverability error of Eq.~\eqref{eq:recoverability} remained below \(7.48\%\). All three error measures increased monotonically as the physical plane moved farther from its focal reference, consistent with the trend in PDF-overlap similarity reported above, and reached their largest values at RADIAL WALL. This behavior provides a quantitative recoverability limit: single-observation reconstruction remains close to the reference distribution near the focal plane and degrades progressively as the reconstructed plane moves away from focus.

The resulting recoverability limit is specific to the present optical configuration, bubble population, segmentation procedure, and inverse model. It therefore represents a quantified operating range specific to the current system.

The inverse framework therefore links DNS-resolved supervision, bubble-instance extraction, focus-conditioned multitask regression, and three-dimensional population assembly within a common reconstruction procedure. Broad-depth controlled-density training extends the supervised focal-plane-relative depth range while preserving correspondence between rendered optical evidence and DNS labels. The framework is evaluated at two complementary levels, reported above together with their respective methodologies: closed-loop DNS validation provided direct bubble-level assessment against known three-dimensional positions and diameters, and experimental validation evaluated depth-resolved bubble-size distributions, \(D_{32}\), \(D_{99.8}\), and recoverability with increasing focal-plane distance. The following section interprets these findings and situates them relative to to existing single- and multi-camera
bubble-imaging methods.

\section{Discussion}
\label{sec:discussion}
The principal scientific finding of this study is that a validated forward optical model, combined with DNS-resolved physical supervision and focus-conditioned multitask regression, resolves the depth--diameter ambiguity inherent to single-camera shadowgraphy to a degree that has previously required stereoscopic or multi-camera imaging. This conclusion rests on three internally consistent pieces of evidence: the forward ray tracer reproduces experimental shadowgraphic signatures to within a few percent (Sec.~\ref{sec:forward_framework}); the inverse model preserves bubble-level depth and diameter information through the complete closed-loop DNS-to-image-to-reconstruction cycle (Sec.~\ref{subsec:closed_loop_validation}); and the same trained model, applied without modification to experimental images, reproduces depth-resolved bubble-size statistics that agree with independent experimental references (Sec.~\ref{subsec:experimental_transfer}, Sec.~\ref{subsec:distribution_validation}). The remainder of this section interprets these results physically, situates them relative to existing bubble-imaging methods, discusses the generalizability and limitations of the present framework, and outlines future extensions.

\subsection{Physical interpretation of the learned reconstruction}
\label{subsec:physical_interpretation}

Explicit focal-plane conditioning (Eq.~\eqref{eq:focus_conditioned_representation}) is central to resolving the depth--diameter ambiguity. Without it, a given combination of apparent extent, edge sharpness, and halo structure is compatible with many absolute depths, since the same local optical appearance can arise at different physical depths simply because the focal-plane setting differs between rendered or experimental frames. Supplying the known focal-plane coordinate as an auxiliary input removes this source of aliasing and allows the network to specialize in inferring only the residual, focal-plane-relative displacement that is actually encoded in the local image evidence, consistent with the depth-from-defocus principle that image sharpness and blur vary systematically with distance from focus~\citep{pentland1987newsense}.

Joint prediction of depth and diameter (Sec.~\ref{subsec:multitask_inverse_model}) is physically appropriate because apparent bubble extent, edge attenuation, and halo development are not separable functions of diameter and depth individually; both a larger physical diameter and a smaller focal-plane displacement can increase apparent size, and both greater displacement and larger diameter can extend halo structure. A shared encoder lets the network learn this coupling directly, consistent with the general expectation that correlated regression targets sharing the same underlying evidence benefit from a shared representation~\citep{ruder2017overview}.

Broad-depth controlled-density training (Sec.~\ref{subsec:broad_depth_learning}) is what extends this behavior across the full working depth range. Because rendering the complete DNS population over a broad depth interval would let strong near-focus signatures dominate and obscure weak far-defocus evidence, the stratified, density-controlled sampling used here is necessary to expose the network to far-defocus examples during training; the resulting normalized depth accuracy of \(\sim97.6\%\) on the held-out synthetic test set indicates that this strategy succeeds in making distant, weakly evidenced bubbles learnable.

The experimental transfer results support a specific interpretation of where residual error originates. Because the ray tracer is independently validated against experimental optical signatures and the inverse model is independently validated in closed loop against known DNS states, the systematic decrease in PDF-overlap similarity from \(\sim86\%\) at the focal reference plane to \(\sim68\%\) at RADIAL WALL (Sec.~\ref{subsec:distribution_validation}) is most consistent with the same physical mechanism identified in the closed-loop analysis, namely the weakening of defocus evidence with distance from focus. This distinction matters because it indicates that the recoverability limit reported here is a property of the single-camera optical configuration and the achievable signal-to-noise ratio of defocus cues, and is not primarily an artifact of training on simulated images.

The composite recoverability error (Eq.~\eqref{eq:recoverability}) formalizes this limit: remaining below \(7.7\%\) at the focal reference plane and increasing monotonically with focal-plane distance, it defines a bounded but quantified operating range within which single-observation reconstruction can be trusted for the present optical configuration, bubble population, and inverse model.

\subsection{Comparison with existing bubble-imaging approaches}
\label{subsec:comparison_existing}

Most learning-based bubble-imaging methods address detection, segmentation, or tracking from projected image evidence and report two-dimensional outputs: bounding boxes and shape regression~\citep{haas2020_bubcnn}, detection combined with tracking~\citep{toyama2025_fasterrcnn}, pixel-level mask extraction~\citep{kim2021_universal}, lightweight instance segmentation for industrial inspection~\citep{yang2026_yolo_bubble}, and general computer-vision or deep-learning pipelines for multiphase bubble characterization~\citep{nizovtseva2024_bubble_detection_math,yang2024_gassolid_bubble_dl,jiang2024_ml_bubble_fluidized}. The bubble-evidence localization and instance-extraction stage of the present framework (Sec.~\ref{subsec:instance_extraction}) follows the same region-proposal-then-mask-assignment paradigm as these methods, but the framework as a whole extends beyond two-dimensional detection by adding physically calibrated depth and diameter regression on top of the extracted instances.

Three-dimensional bubble measurement has typically relied on multiple synchronized cameras: deep-learning-assisted three-dimensional tracking of deformable bubble swarms~\citep{Hessenkemper2024_3DBubbleTracking} and temporal information used to improve cross-view association in multi-camera three-dimensional tracking~\citep{wang2025_temporal_3d_bubble_tracking} both achieve three-dimensional trajectories and require the additional cameras, calibration, and synchronization that the present single-camera framework is designed to avoid. The present work is therefore complementary to multi-view methods: it targets configurations in which optical access, cost, or synchronization constraints preclude a multi-camera setup, at the cost of the bounded recoverability range quantified in Sec.~\ref{subsec:distribution_validation}.

Within the present research program, the preliminary depth-aware pipeline of \citet{nayak2026depth} used unsupervised clustering of engineered focus-sensitive features to produce a relative, unitless focus score and a binary in-focus/out-of-focus label. The present framework replaces that engineered, relative pipeline with DNS-resolved physical supervision, an experimentally validated ray-tracing forward model, and end-to-end multitask regression, extending the earlier qualitative focus ordering to signed physical depth and physical-equivalent diameter validated against both known DNS states and independent experimental population statistics. More broadly, the strategy of validating a physically based forward renderer against real measurements and then using it to supply supervision for a learned inverse model parallels recent demonstrations in other optical inverse problems, including ray-tracing image simulation of transparent refractive objects~\citep{kalita2026_ray_tracing_transparent} and machine-learning-based inversion of pendant-drop images for surface tension~\citep{kratz2020_pendant_drop_ml}, suggesting that the present approach is an instance of a more general, physically grounded simulation-to-inversion strategy.

\subsection{Limitations and generalizability}
\label{subsec:limitations_generalizability}

The present results should be interpreted within the validated physical and optical range of the study. The validation framework tests two forms of generalization. Timestamp-disjoint partitioning (Sec.~\ref{subsec:synthetic_supervision}) evaluates performance on DNS turbulence states that are not included in training, while experimental transfer (Sec.~\ref{subsec:experimental_transfer}) evaluates the transition from simulated shadowgraph images to experimentally acquired images. In both cases, the regression targets are expressed directly in physical units as signed focal-plane-relative depth and volume-equivalent diameter. This preserves the physical meaning of the predictions and keeps the inverse model tied to the calibrated imaging geometry.

The current validation does not establish generalization across all flow and imaging conditions. The DNS database spans two void fractions (\(0.5\%\) and \(2\%\)) and one air--water density and viscosity ratio. Changes in void fraction, gas--liquid properties, or Weber-number regime can alter bubble deformation, spatial organization, and overlap statistics, and these conditions have not been evaluated here. The optical model and inverse network are also calibrated for the camera, magnification, illumination, focal-plane positions, and exposure settings described in Sec.~\ref{sec:forward_framework}. A different imaging configuration would change the relationship between defocus and signed depth through quantities such as numerical aperture and depth of field and would therefore require corresponding optical recalibration and, where needed, new rendered training data.

Additional limitations arise from the forward model and the inverse reconstruction sequence. Although the ray tracer reproduces the principal experimental shadowgraphic signatures to within a few percent, it does not include every optical effect present in the physical experiment. The difference between the closed-loop diameter accuracy (\(\sim90.29\%\)) and the highest experimental distributional similarity (\(\sim86.37\%\) at the focal reference plane) therefore reflects the combined influence of forward-model mismatch, experimental variability, localization errors, instance-separation errors, and regression uncertainty. The present data do not permit these contributions to be isolated quantitatively.

Bubble overlap remains another source of uncertainty. The controlled-density training strategy reduces severe occlusion during learning, but overlap is still present in both rendered and experimental images. At higher local bubble concentrations, neighboring instances may merge or become partially obscured during localization and instance extraction. Errors introduced at this stage propagate into the crop representation and subsequently affect the predicted depth and diameter. The closed-loop validation in Sec.~\ref{subsec:closed_loop_validation} includes this compounded behavior because the complete reconstruction pipeline is evaluated from rendered image to recovered bubble population.

Recoverability also decreases with increasing distance from the focal plane. At large defocus, edge contrast, halo structure, and other depth-sensitive image features progressively weaken and approach the experimental noise level. The degradation reported in Sec.~\ref{subsec:distribution_validation} therefore reflects a reduction in the available optical information at large focal-plane displacement. This behavior defines a practical depth range over which individual bubbles can be reconstructed with useful confidence.

The predicted physical diameter introduces an additional constraint. Diameter is represented by a volume-equivalent scalar even though the DNS-derived renderings retain the underlying non-spherical interface geometry described in Sec.~\ref{sec:dns_dataset}. Consequently, deformation information contained in the projected bubble shape is compressed into a single diameter target. For strongly deformed bubbles, similar projected signatures may correspond to different three-dimensional shapes and volume-equivalent diameters, which introduces uncertainty into the inverse mapping.

Experimental validation is further constrained by the absence of bubble-resolved three-dimensional ground truth. Direct correspondence between each experimentally detected bubble and its true depth and diameter is therefore unavailable. Experimental performance is assessed through depth-resolved population statistics and characteristic diameter measures, including full-distribution PDF agreement, \(D_{32}\), and \(D_{99.8}\). These metrics provide a population-level assessment of reconstruction quality but do not replace bubble-to-bubble three-dimensional validation.

Taken together, these considerations define the current operating envelope of the method. Generalization has been demonstrated across unseen DNS timestamps and from rendered to experimental shadowgraph images within the calibrated flow and imaging conditions. Extension to substantially different void fractions, fluid-property ratios, deformation regimes, optical configurations, or focal ranges remains to be established as future work through additional DNS datasets, optical recalibration, and experimental validation.

\subsection{Future work}
\label{subsec:future_work}

Several extensions follow naturally from the present framework. Temporal tracking of reconstructed bubbles across consecutive frames could exploit the sample-identity bookkeeping already used for closed-loop validation (Sec.~\ref{subsec:closed_loop_validation}) to regularize depth and diameter estimates over time and to extract velocity and acceleration statistics. Sparse multi-view imaging, using a second camera at a coarser sampling rate than a full stereoscopic rig, could directly target the recoverability degradation at large focal-plane distance without incurring the full calibration and synchronization cost of dense multi-camera systems. Uncertainty quantification, for example through learned heteroscedastic weighting of the multitask objective~\citep{kendall2018multitask} in place of the fixed weights used in Eq.~\eqref{eq:multitask_loss}, would allow per-bubble confidence to propagate into population-level statistics as well as aggregate error metrics. Explicit domain-adaptation techniques, applied in addition to forward-model validation, could further reduce the residual gap observed at the more challenging measurement planes. Physics-informed loss terms that directly encode the depth-from-defocus relation~\citep{pentland1987newsense}, in addition to focal-plane conditioning, could provide additional inductive bias, particularly in the far-defocus regime identified as a limitation above. Predicting a shape or aspect-ratio descriptor alongside the volume-equivalent diameter would recover some of the non-sphericity information currently discarded by the scalar diameter target. Finally, extending the DNS database to higher void fractions than the \(0.5\%\) and \(2\%\) cases studied here would test the framework under heavier bubble-to-bubble overlap and would help determine whether instance extraction or depth--diameter regression becomes the limiting stage as bubble density increases.

\section{Conclusions}
\label{sec:conclusions}
This work addresses a central limitation of single-camera shadowgraphy: a two-dimensional bubble image does not uniquely reveal either physical diameter or three-dimensional position because both quantities affect the observed optical signature. The approach taken here was to supply the inverse problem with physically grounded supervision rather than rely on purely image-based correlations. DNS-resolved bubble fields were converted into camera-like shadowgraphs using an experimentally validated ray-tracing model, individual bubble observations were isolated from full frames, and a focus-conditioned multitask network was trained to recover focal-plane-relative depth and physical-equivalent diameter.

The results show that the optical information required for three-dimensional reconstruction is retained in a single shadowgraphic view when the forward imaging physics and focal geometry are incorporated into the learning framework. The reconstruction remained accurate on DNS states excluded from training, preserved bubble geometry through the complete three-dimensional-to-image-to-three-dimensional cycle, and transferred to experimental shadowgraphs without retraining. Experimental comparisons further showed that recovery is strongest near the focal reference plane and degrades progressively with increasing defocus. This behavior is physically meaningful and provides a practical measure of the depth range over which reliable population statistics can be extracted from a given optical configuration.

The main outcome is therefore not only a bubble-regression model, but a general strategy for converting high-fidelity simulation into experimentally usable inverse-imaging supervision. By linking resolved multiphase physics, validated image formation, and learned inversion, the framework enables depth-resolved bubble-size measurements from a single optical view while retaining a direct connection between reconstructed quantities and their physical origin. This reduces the dependence on stereoscopic access and offers a practical route toward three-dimensional characterization of bubbly flows in experiments and multiphase processes where multi-camera measurements are difficult or impossible.

\backmatter

\section*{Acknowledgements}
This work was supported in part by the Advanced Research Projects Agency--Energy, U.S. Department of Energy (DOE) under Award Number DE-AR0001587, and Office of Critical Minerals and Energy Innovation (Former EERE), US DOE under Award Number DE-EE0009396. \\S.S.J. acknowledges support from the donors of the ACS Petroleum Research Fund through Doctoral New Investigator Grant 69196-DNI9 and the NSF CAREER award with grant number GR00037347. 
The authors also acknowledge computing resources provided through the U.S. Department of Energy 2024 and 2025 ALCC awards (TUR147 and BubbleLaden). The Argonne Leadership Computing Facility at Argonne National Laboratory is supported by the Office of Science of the U.S. Department of Energy under Contract No. DE-AC02-06CH11357. The Oak Ridge Leadership Computing Facility at Oak Ridge National Laboratory is supported by the Office of Science of the U.S. Department of Energy under Contract No. DE-AC05-00OR22725.  \\
Chaitanya S. Nayak and Shivam Prajapati received funding from the Renewable Bioproducts Institute (RBI) Graduate Fellowship at the Georgia Institute of Technology. 

\section*{Declarations}

\bmhead{Conflict of interest} The authors declare no competing interests.

\vspace{0.5em}

\bmhead{Ethics approval} Not applicable.

\vspace{0.5em}

\bmhead{Consent to participate} Not applicable.

\vspace{0.5em}

\bmhead{Consent for publication} Not applicable.

\vspace{0.5em}

\bmhead{Availability of data and materials} The data are available from the corresponding author upon reasonable request. Data are located in controlled access data storage at Georgia Institute of Technology.

\vspace{0.5em}

\bmhead{Authors' contributions}

C.S.N. and F.M. contributed equally to this work.

C.S.N.: Study conception, deep-learning methodology, architecture development, visualization, and manuscript preparation.

F.M.: Study conception, ray-tracing methodology, optical-model development, visualization, and manuscript preparation. 

V.K.: Study conception, experimental methodology, experimental data acquisition, dataset preparation, validation, and resources.

S.P.: Study conception, literature review, data interpretation, and manuscript preparation.

S.S.J.: Study conception, methodological guidance, supervision, validation, funding acquisition, and manuscript preparation.

C.K.A.: Study conception, methodological guidance, supervision, project administration, resources, funding acquisition, and manuscript preparation.

All authors reviewed and approved the final manuscript.

\bibliography{sn-bibliography}

\end{document}